\documentclass[%
 reprint,
superscriptaddress,
nofootinbib,
 amsmath,amssymb,
  prd,
]{revtex4-1}

\usepackage{latexsym, amsmath, amssymb,amsthm,amsopn,amsfonts, amsbsy, graphicx, epstopdf, multirow}
\usepackage{breqn}
\usepackage{color}
\usepackage{ulem}
\usepackage{graphicx}
\usepackage{cleveref}

\def\b2otimes{\boldsymbol{\bar\otimes}}

\newcommand{\noop}[1]{}

\def\simlt{\lower.5ex\hbox{$\; \buildrel < \over \sim \;$}}
\def\simgt{\lower.5ex\hbox{$\; \buildrel > \over \sim \;$}}
\def\simgtalt{\lower.5ex\hbox{$\buildrel > \over \sim \;$}}

\def\b#1{{\bar{#1}}}

\def\bd#1{{\bf #1}}
\def\l#1{\left #1}
\def\r#1{\right #1}

\makeatletter
\let\cat@comma@active\@empty
\makeatother

\begin{document}
\title{Characterizing bias on large scale CMB $B$-modes after Galactic foregrounds cleaning}

\author{Josquin Errard}
\email{josquin@apc.in2p3.fr}
%\affiliation{AstroParticule et Cosmologie, Univ Paris Diderot, CNRS/IN2P3,CEA/Irfu, Obs de Paris, Sorbonne Paris Cit\'e, France}
\affiliation{Universit\'e de Paris, CNRS, Astroparticule et Cosmologie, F-75013 Paris, France}
\author{Radek Stompor}
\email{radek.stompor@in2p3.fr}
%\affiliation{AstroParticule et Cosmologie, Univ Paris Diderot, CNRS/IN2P3,CEA/Irfu, Obs de Paris, Sorbonne Paris Cit\'e, France}
\affiliation{Universit\'e de Paris, CNRS, Astroparticule et Cosmologie, F-75013 Paris, France}

\date\today

\begin{abstract}
We study the performance of a typical near-future full sky CMB space mission, aiming at the characterization of the large scale $B$-modes polarization anisotropies with precision on $r\sim 0.001$, after a map-based parametric cleaning of galactic dust and synchrotron, and in the case of spatially varying astrophysical spectral emission laws. Ignoring the spatial variability of the spectral emission laws may result in a bias on $r$ as high as $\mathcal{O}(0.01)$ for realistic models of the variability. 
However, we show that the component separation formalism can be extended to suppress this bias efficiently. We demonstrate this within the context of the semianalytic formalism of~\citep{Stompor2016}, which we generalize to such cases and use it to propagate the foreground residuals to a cosmological likelihood on tensor-to-scalar ratio, $r$. In particular, we investigate the effects due to introducing extra, independent sets of scaling parameters for different sky areas and including additional scaling parameters per sky area. We then show that the residuals resulting in such cases can be efficiently described with help of extra terms introduced in a model of the covariance of the component-separated CMB maps and which lead to suppression of the bias on $r$ down to the level lower than the expected statistical uncertainty. We discuss how these additional terms can be constructed self-consistently from the available data.
\end{abstract}

\maketitle
%%%%%%%%%%%%%%%%%%%%%%%%%%%%%%%%%%%%
\section{Introduction}

The characterization of CMB polarization is in a prosperous period, with many ground-based projects already observing (among which POLARBEAR~\cite{2017ApJ...848..121P}, BICEP-Keck~\cite{2018arXiv180702199T}, SPTPol~\cite{2018ApJ...852...97H}, ACTPol~\cite{2017JCAP...06..031L}, BICEP 3~\cite{2016SPIE.9914E..0SG}, Simons Array~\cite{2018AAS...23135602B}, Advanced-ACTPol~\cite{2018JLTP..tmp..159C}, SPT-3G~\cite{2018arXiv180900032S}) or about to observe (e.g. Simons Observatory~\cite{Simons_Observatory}, BICEP Array~\cite{2018arXiv180800568H}, CMB-S4~\cite{2016arXiv161002743A}), along with proposed space projects such as LiteBIRD~\cite{2016JLTP..184..824M}, PIXIE~\cite{2016SPIE.9904E..0WK} or PICO~\cite{PICO}. Although their science cases are generally broad, the characterization of the $B$-modes polarization patterns on the largest angular scales is one of the most exciting goals: if detected, it would help us understanding the inflationary mechanism, and open up an observational window onto the physics at the highest energies~\cite{1998cosm.work..355L}. If not detected, we would be able to discriminate entire class of inflationary models~\cite{Martin_2014a,Martin_2014b}. 
Yet, reaching this tiny signal requires an exceptional instrumental sensitivity and exquisite control of the noise on the largest angular scales as well as systematic effects. We study in this paper polarized galactic foregrounds, constituting one of the most important contaminants, which, if not taken into account in the analysis, can produce spurious polarization signal corresponding to values of the tensor-to-scalar ratio, $r$, as high as $0.01$ in the cleanest frequency bands as well as in the cleanest parts of the sky~\cite{Krach2016}.

Polarized galactic foregrounds are dominated by dust and synchrotron emissions, which are still only poorly known with the best constraints due to Planck~\cite{Planck_LIV_2018,2018arXiv180706212P,2018arXiv180104945P}. The limited knowledge of spectral emission densities (SEDs) of these signals leaves the possibility of a rather complex sky, even at high galactic latitudes. This view is further corroborated by
several theoretical and observational works~\cite{Vansyngel2017, Levrier_2018, 2018arXiv180201145K}, which indicate that spectral indices, parametrizing synchrotron and dust SEDs, are typically to be expected to vary across the sky. If this is indeed the case, the component separation methods need to estimate these emissions in various parts, or regions, of the sky. 
This will be particularly pertinent in the analysis of (nearly) entire sky data, as expected from future satellites, where neglecting the SEDs variability could lead to a false detection of the tensor-to-scalar ratio $r\approx 0.005-0.01$~\cite{Planck_L_2017,2016AA...594A..25P}.
However, given the projected and limited sensitivity of the future CMB instruments, such a generalization of the foreground cleaning procedure will unavoidably affect the quality of the characterization of the SEDs, consequently, allowing for more of the leaked galactic foregrounds signal in the final, estimated, cleaned CMB map~\cite{Stivoli2010}. This could potentially undermine the feasibility of reaching the scientific target of $r\leq 0.001$ as defined for the future missions, even in the cases when the spatial dependence of the foregrounds spectral indices is known a priori. 
In this paper we generalize the approach of~\cite{Stompor2016} and study foreground residuals arising in such more general applications, showing how to model residuals of a different origin and how to incorporate them in the cosmological likelihood on $r$.

The paper is organized as follows. In section~\ref{sec:method} we describe our component separation method, which is based on a parametric, pixel-based maximum-likelihood approach --- but adapted to deal with the spatial variability of spectral emission laws. We also detail the assumptions regarding the sky simulations, the specifications of a typical space CMB instrument, as well as the implementation of the algorithm. 
We present our results in section~\ref{sec:results}, which are split between two cases: an ideal case where we know \textit{a priori} the regions where spectral indices can be assumed constant, and a more realistic case where we do not have such information. 
Finally, we discuss our results and the potential developments of such method in section~\ref{sec:conclusions}.

%%%%%%%%%%%%%%%%%%%%%%%%%%%%%%%%%%%%
\section{Method}
\label{sec:method}

We adopt the parametric component separation method~\cite{brandt1994,2006ApJ...641..665E} throughout this work and therefore assume that dust and synchrotron spectral emission distributions can be parametrized by a set of parameters denoted $\boldsymbol{\beta}$.
The fact that we let the SEDs to vary between different regions of the sky~\cite{Planck_XI_2014} or even different line-of-sights has two major consequences which have to be accounted for in a high precision analysis: we not only need to allow for different scaling laws in different sky areas, but may have to adopt more complex scaling laws, which appear naturally as a result of averaging the signal over multiple line-of-sight falling into the same sky pixels~\cite{2017MNRAS.472.1195C}. This would also be expected due to averaging over different SEDs along the same line of sight, or even different SEDs due to multiple dust species~\cite{2015ApJ...798...88M,Vansyngel2017}. This latter effect will typically call for more scaling parameters \textit{per sky area} needed to describe such a more complex scaling law, while the former will need for \textit{more sets of the scaling parameters} used to characterize different sky areas. 

One of the attractive features of the pixel-based parametric technique is that it straightforwardly permits assigning different sets of spectral parameters to different sky areas. In the most extreme case, this can correspond to having a single set of spectral parameters for each considered sky pixel. On the other extreme, there will be only one set of spectral parameters for the entire available map. The former case is in principle the most conservative and robust as it would lead to no bias in the results as long as the assumed scaling laws are sufficiently close to the true ones. However, this comes at the cost of as many as $n_{\boldsymbol{\beta}}\,n_{\rm pix}$ extra degrees of freedom and is therefore bound to boost up significantly the statistical uncertainty of the estimates. In contrast, assuming a single set of spectral indices for the entire map, while clearly beneficial from the point of view of the statistical errors, can be potentially detrimental as far as the bias is concerned. Given what we know about the polarized foregrounds, this latter occurrence is indeed likely to be the case and we therefore may need to consider intermediate options which trade the systematic for the statistical uncertainty, in principle allowing to achieve the smallest statistical errors while keeping the systematic ones negligible.

Therefore, in order to account for the spatial variability of spectral indices, we explore the possibility of dividing the full observed sky into a set of disjoint sky regions, and assigning a different set of spectral indices to each of those. We then fit independently for each of the sets of indices. For simplicity and definiteness, the regions are defined to follow a \textsc{Healpix} grid~\cite{healpix} with a resolution set by the \textsc{Healpix} parameter $\bar{n}^\beta_{\rm side}$. 
Similarly, the true sky scaling parameters are assumed to vary over another \textsc{Healpix} grid with a resolution defined by $n^{\beta}_{\rm side}$. In this work, for fixed value of $n^{\beta}_{\rm side}$, we study settings with different values of the $\bar{n}^\beta_{\rm side}$ parameters,
$0\leq\{\bar{n}^\beta_{\rm side},n^\beta_{\rm side}\}\leq n^{\rm map}_{\rm side}$,  thus exploring intermediate cases for which the trade-off between the statistical and systematic errors is different. While these assumptions are clearly oversimplifying the actual circumstances, they allow us to investigate and clarify the impact of different factors in a controlled manner.

\begin{figure*}
	\centering
		\includegraphics[width=\textwidth]{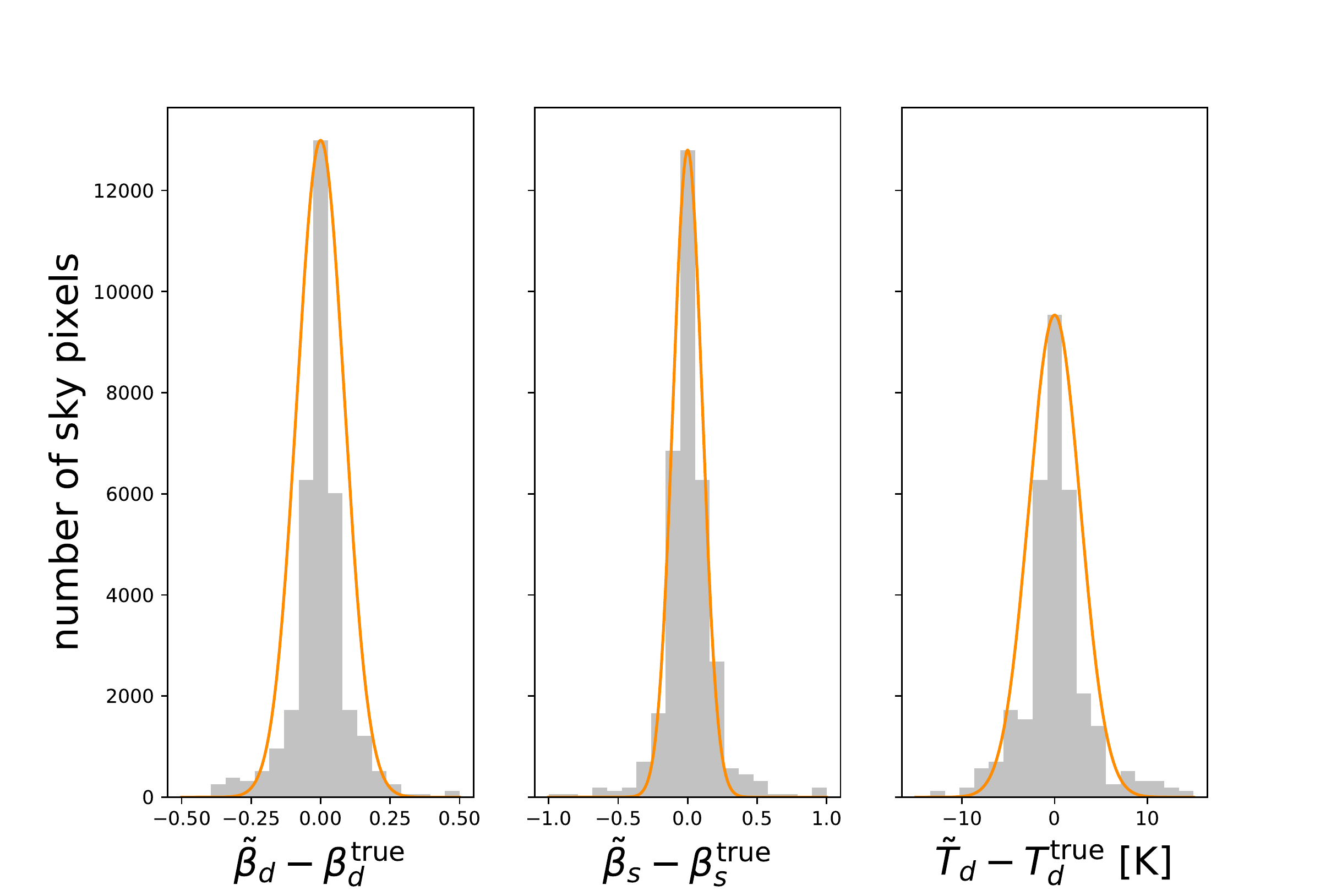}
	\caption{Histograms of the recovered spectral parameters $\boldsymbol{\beta} \in \{\beta_d, \beta_s, T_d\}$, in the case of $n_{\rm side}^\beta = \bar{n}_{\rm side}^\beta = 8$ (see section~\ref{ssec:ideal_case} for more details). These are estimated through the optimization of the spectral likelihood, Eq.~\ref{eq:estimation_of_beta}, for a given realization of CMB and instrumental, white noise. The orange curves are Gaussian, centered on zero, with standard deviations given by the diagonal elements of $\boldsymbol{\Sigma}$, estimated with Eq.~\ref{eqn:secDervSat}, averaged over sky patches.}
	\label{fig:histogram_delta_beta}
\end{figure*}

\subsection{Rendering of parametric maximum-likelihood component separation}

Similarly to e.g.~\citep{Stompor2009}, for each sky pixel $p$, we model the data as,
\begin{eqnarray}
	\centering
		d_p^{\nu,\rm obs} = \mathbf{A}^\nu(\boldsymbol{\beta}_{(\mathcal{P})})s_p + n^{\nu,\rm obs}_p,
	\label{eq:data_modeling}
\end{eqnarray}
where,\\
$\bullet$ \ $d_p^{\nu,\rm obs}$ is a vector containing the observed Stokes parameters, $\{Q,U\}$, at a frequency $\nu$;\\
$\bullet$ \ 
$s_p$ is the sky signal containing the $\{Q,U\}$ amplitudes of the real sky components, cosmological and astrophysical, namely 
\begin{eqnarray}
	\centering
		s_p \equiv \{Q_p^{\rm CMB}, U_p^{\rm CMB}, Q_p^{\rm dust}, U_p^{\rm dust}, Q_p^{\rm sync}, U_p^{\rm sync}\},
	\label{eq:s_p_definition}
\end{eqnarray}
if we assume that the polarized sky is composed of CMB, dust and synchrotron,\\
$\bullet$ \ $n_p^{\nu,\rm obs}$ is the frequency-dependent instrumental noise, and,
\\
$\bullet$ \ $\mathbf{A}^\nu$ is the so-called mixing matrix, parametrized by a set of spectral indices, $\boldsymbol{\beta}$, which can be different for each sky patch, $\mathcal{P}$, ranging from the full sky to individual pixels. 
We distinguish between the true spectral indices $\boldsymbol{\beta}^{\rm true}$, used to simulate the observed frequency maps following Eq.~\ref{eq:data_modeling}, and the estimated spectral indices, $\tilde{\boldsymbol{\beta}}$, which are recovered from the mock data sets as discussed in the following. The latter not only can be assumed to vary over the sky differently than the former but can indeed have a different physical interpretation.
Specifically, as mentioned earlier, we assume that $\boldsymbol{\beta}^{\rm true}$ is uniquely defined for each  \textsc{Healpix} pixel with a resolution $n^\beta_{\rm side}$. This can be different from the \textsc{Healpix} grid with a resolution $\bar{n}^\beta_{\rm side}$, used to define the sky patches, $\mathcal{P}$, and assumed during the component separation. For each of these regions,  we estimate a different set of spectral parameters, $\tilde{\boldsymbol{\beta}}$.\\

For a given set of observed frequency maps, $d_p^{\nu,\rm obs}$, and the knowledge of the noise covariance, $\mathbf{N}\equiv\mathbf{N}^{\nu\nu'}_{pp'}\equiv \langle n^{\nu}_{p}{n^{\nu'}_{p'}}^T\rangle$, it is possible to estimate the set of spectral parameters $\boldsymbol{\beta}$ and hence the scaling law of each sky component:
\begin{eqnarray}
	\centering
		\tilde{\boldsymbol{\beta}} \equiv {\rm argmax}\left( \mathcal{L}_{\rm spec}(\boldsymbol{\beta}) \right)
	\label{eq:estimation_of_beta}
\end{eqnarray}
where the spectral likelihood, $\mathcal{L}_{\rm spec}$, is defined as,~\cite{Stompor2009},
\begin{eqnarray}
	\centering
		&&\hspace{-0.5cm}-2\log (\mathcal{L}_{\rm spec}(\boldsymbol{\beta})) \equiv \nonumber\\&&\hspace{1.0cm}\left( \mathbf{A}^T\mathbf{N}^{-1}d \right)^T \left( \mathbf{A}^T\mathbf{N}^{-1}\mathbf{A} \right)^{-1} \left( \mathbf{A}^T\mathbf{N}^{-1}d \right).
	\label{eq:spectral_likelihood}
\end{eqnarray}
This likelihood can be either obtained via maximization of the original likelihood over the sky signal estimates or marginalization over them~\citep{Stompor2009}. In this latter case, an additional term proportional to the logarithm of the determinant of the sky signal covariance would appear on the right hand side leading to biases of the estimated spectral parameters~\citep{Stompor2009}. This can be corrected for by adapting a Jeffrey-like prior on spectral parameters~\citep{Alonso2016}.
%As also derived in the appendices of~\citet{Alonso2016}, one will note that we forget about a determinant term in the equation above.

Eq.~\ref{eq:spectral_likelihood}  assumes implicit sums over sky pixels $p$ and frequencies $\nu$ so that $-2\log (\mathcal{L}_{\rm spec}(\boldsymbol{\beta}))$ is a scalar. 
It can be used not only to estimate the best-fit value of $\boldsymbol{\beta}$, $\tilde{\boldsymbol{\beta}}$, Eq.~\ref{eq:estimation_of_beta}, but also to approximate its uncertainty by the curvature of the spectral likelihood at its peak, i.e., 
\begin{eqnarray}
	\centering
		\sigma(\beta_i) &\approx& \mathbf{\Sigma}^{1/2}_{\left(\beta_i, \beta_i\right)} \label{eq:sigma_beta_def}\\
		{\rm with}\hspace{0.5cm} \left[\mathbf{\Sigma}^{-1}\right]_{\left( \beta_i,\beta_j\right)}&\equiv& \left.\frac{\partial^2 \mathcal{L}}{\partial \beta_i \partial \beta_j}\right|_{\boldsymbol{\tilde\beta}}, 
		\label{eq:Sigma_def}
\end{eqnarray}
where, as in any Fisher approach, we first need to invert the Hessian defined by Eq.~\ref{eq:Sigma_def} before taking the square root of the $(\beta,\beta)$ element of the inverse matrix, Eq.~\ref{eq:sigma_beta_def}, so that the estimated uncertainty of the parameter $\beta$, $\sigma(\beta)$, correctly incorporates the effect of marginalization over all other spectral parameters. \citet{Errard2011} proposed a semianalytical expression for the average curvature of the spectral likelihood, Eq.~\ref{eq:Sigma_def}, which reads,
\begin{eqnarray}
&&\l[\bd{\Sigma}_{\rm analytic}^{-1}\r]_{\left(\beta_i,\beta_j\right)} \equiv \left\langle \left.\frac{\partial^2 \mathcal{L}}{\partial \beta_i \partial \beta_j}\right|_{\boldsymbol{\tilde\beta}} \right\rangle_{\rm noise} \simeq 
\nonumber\\
&& -{\rm tr} \Biggl( \Biggl[ \left.\frac{\partial \bd{A}}{\partial \beta_i}\right|_{\boldsymbol{\tilde\beta}}^T \bd{N}^{-1} \bd{A} \l(\bd{A}^T\bd{N}^{-1}\bd{A}\r)^{-1} \bd{A}^T\bd{N}^{-1} \left.\frac{\partial \bd{A}}{\partial \beta_j}\right|_{\boldsymbol{\tilde\beta}} 
\nonumber\\ 
&& \ \ \ \ \ \ \ \ \ \ \ - \left.\frac{\partial \bd{A}}{\partial \beta_i}\right|_{\boldsymbol{\tilde\beta}}^T \bd{N}^{-1} \left.\frac{\partial \bd{A}}{\partial \beta_j}\right|_{\boldsymbol{\tilde\beta}} \Biggr] \sum_p \bd{s}_p \bd{s}_p^T \Biggr), 
\label{eqn:secDervSat}
\end{eqnarray}
In this work, we take $\bd{\Sigma}_{\rm analytic}$ given by Eq.~\ref{eqn:secDervSat} as the approximation of the error bar on $\boldsymbol{\beta}$ (we notice that it is in general numerically more stable than computing the Hessian at the peak of the spectral likelihood).
Given the estimated spectral parameters, $\tilde{\boldsymbol{\beta}}$, (Eq.~\ref{eq:estimation_of_beta}) and therefore the estimated mixing matrix $\tilde{\mathbf{A}} \equiv \mathbf{A}(\tilde{\boldsymbol{\beta}})$, the estimate of the sky signal, $\tilde{s}$, is given by:
\begin{eqnarray}
	\centering
		\tilde{s} &=&  \left(\tilde{\mathbf{A}}^T\mathbf{N}^{-1}\tilde{\mathbf{A}} \right)^{-1} \tilde{\mathbf{A}}^T\mathbf{N}^{-1}d \equiv \mathbf{W}d
	\label{eq:sky_signal_estimate}
\end{eqnarray}
where $d$ are the observed frequency maps introduced in Eq.~\ref{eq:data_modeling}. 
The noiseless foregrounds residuals, $\Delta^{\rm CMB}$, are defined as the difference between the recovered and true CMB sky corrected for the presence of the noise,
\begin{eqnarray}
	\centering
		\Delta^{\rm CMB} \equiv  \tilde{s}^{\rm CMB} - s^{\rm CMB} - n^{\rm CMB}
	\label{eq:tot_fgs_residuals}
\end{eqnarray}
where $n^{\rm CMB}$ is the actual noise in the final CMB map as propagated from the multifrequency data,
\begin{eqnarray}
	\centering
      n^{\rm CMB} \equiv \left[ \mathbf{W} n\right]_{\rm CMB}.
	\label{eq:noise_after_comp_sep}
\end{eqnarray}
Whenever $\langle\boldsymbol{\tilde{\beta}}\rangle = \boldsymbol{\beta}_{true}$ we can develop the foreground residuals to first order in $\boldsymbol{\delta \beta}\equiv \boldsymbol{\tilde{\beta}} - \boldsymbol{\beta}^{\rm true}\ll \boldsymbol{\beta}^{\rm true}$, and write~\citep{Stivoli2010}:
\begin{eqnarray}
	\centering
		\Delta^{\rm CMB} = \Delta^{\rm CMB}_{\rm stat} \approx \sum_{j,k} \delta \beta_j \alpha_k^{j}(\tilde{\beta}) s_p^{(k)},
	\label{eq:stat_fgs_residuals}
\end{eqnarray}
what defines the statistical part of the residuals, i.e., those generated by the statistical uncertainty in a determination of the spectral parameters from noisy data.
The $\alpha$ elements in Eq.~\ref{eq:stat_fgs_residuals} are given by
\begin{eqnarray}
	\centering
		\alpha_k^{j}(\tilde{\beta}) &\equiv& -\left[\left(\mathbf{A}^T\mathbf{N^{-1}}\mathbf{A}\right)^{-1} \mathbf{A}^T\mathbf{N^{-1}}  \left.\frac{\partial \mathbf{A}}{\partial \beta_{k}}\right|_{\tilde{\beta}_k} \right]_{\left(\rm CMB,\, j\right)}
	\label{eq:alpha_def}
\end{eqnarray}
where $(CMB, j)$ denotes a relevant element of the 2D matrix and each $\delta \beta_j$ element in Eq.~\ref{eq:stat_fgs_residuals} follows approximately
a Gaussian distribution with mean $0$ and a standard deviation given by $\sigma(\tilde{\beta})$, as approximated in Eq.~\ref{eq:sigma_beta_def}. 
As described in~\cite{Stivoli2010,Errard2011,Stompor2016}, nonzero uncertainties on the recovered spectral parameters i.e. $\sigma(\tilde{\beta}_j) \neq 0$ would necessarily lead to the nonvanishing power of the statistical foregrounds residuals in the clean CMB map, i.e. $ \Big\langle \left(\Delta^{\rm CMB}\right)^2 \Big\rangle \neq 0$.

More generally, for instance, when the true spectral parameters, $\boldsymbol{\beta}^{\rm true}$, are spatially varying, but our adopted model assumes a single set of $\tilde{\boldsymbol{\beta}}$ over the entire sky, $\langle\boldsymbol{\tilde{\beta}}\rangle \ne \boldsymbol{\beta}_{true}$ and the foreground residuals will be composed of two terms, the statistical residuals due to the scatter of actual recovered values of $\boldsymbol{\tilde{\beta}}$ around their average, $\langle\boldsymbol{\tilde{\beta}}\rangle$, and systematic residuals due the fact that $\langle\boldsymbol{\tilde{\beta}}\rangle \ne \boldsymbol{\beta}_{true}$, i.e.,
\begin{eqnarray}
	\centering
		\Delta^{\rm CMB} = \Delta_{\rm stat}^{\rm CMB} + \Delta_{\rm sys}^{\rm CMB}.
        \label{eq:resBreakDown}
\end{eqnarray}

In such circumstances, by amending our modeling of foregrounds SEDs by adding more $\boldsymbol{\beta}$ parameters we can potentially trade the systematic for statistical residuals.

\subsection{Toward a ``multipatch'' approach}
\label{ssec:generalization_multipatch}
The approach we consider here is arguably the most straightforward extension of the standard parametric component separation technique that can accommodate spatial variability of the foreground SEDs. We refer to it as a multipatch approach as it invokes some partition of the observed sky in a set of disjoint sky regions called hereafter ``patches'', and assumes that a different set of the spectral parameters, $\boldsymbol{\beta}$, is assigned to each of them. All these parameters have to be determined as part of the component separation procedure before the cleaned CMB map can be produced. As our goal is to study the effects of introducing these additional degrees of freedom on the quality of the separation procedure, we leave aside practical issues such as how in practice such patches should be defined (see, however, Section~\ref{sec:conclusions}). \\

\begin{figure}
	\centering
		\includegraphics[width=\columnwidth]{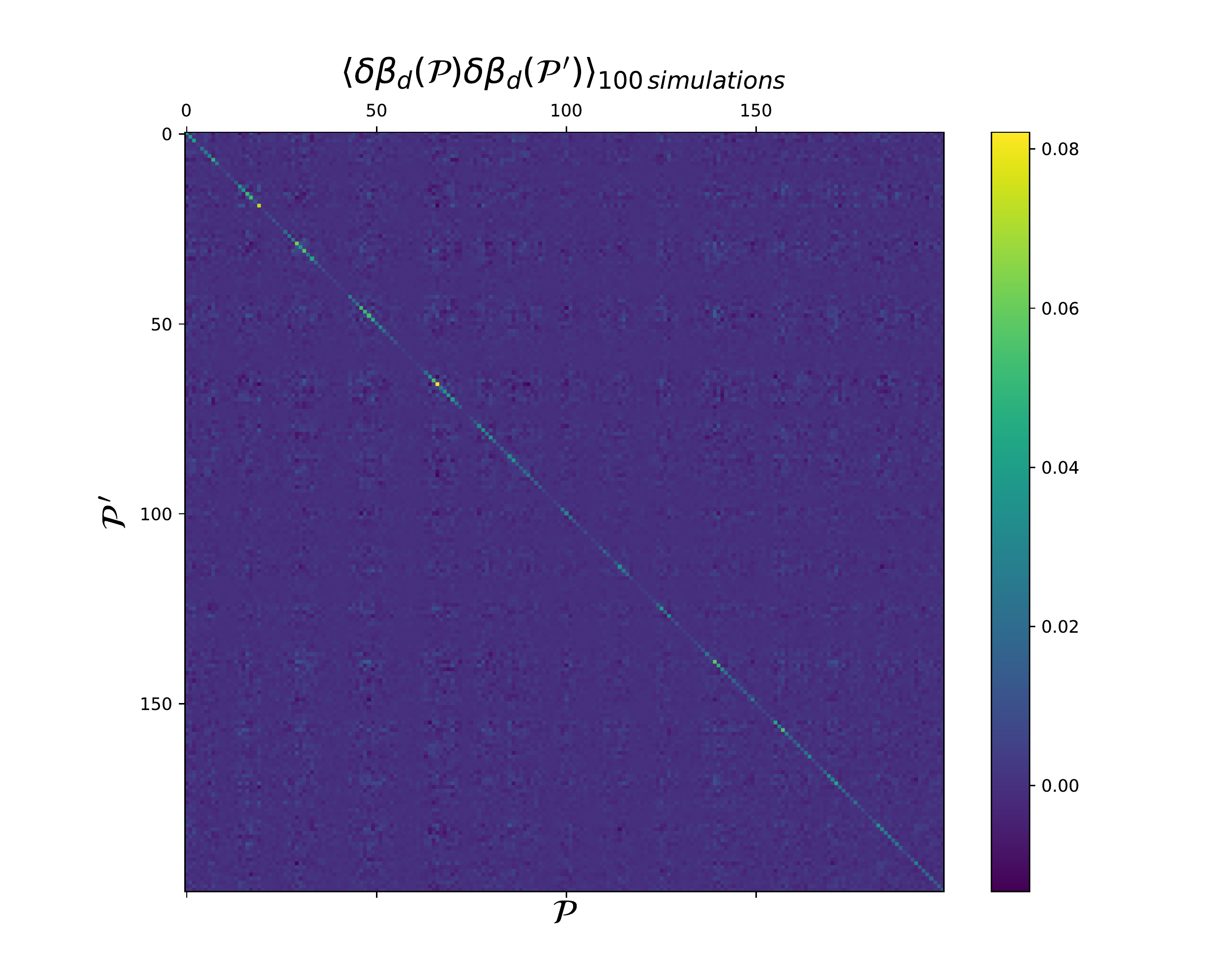}
	\caption{$\left\langle \delta\beta_d(\mathcal{P}) \delta\beta_d(\mathcal{P}^\prime) \right\rangle_{\rm noise}$, as a function of $\mathcal{P}$ and $\mathcal{P}^\prime$, averaged over 100 simulations of noise. This matrix is diagonal dominated, as suggested by Eqs.~\ref{eq:diagonal_delta_beta_square} and~\ref{eq:off_diagonal_delta_beta_square}. The variability of the amplitude of the diagonal elements is due to the variability of the foregrounds amplitude across the sky, since $\delta\beta\sim\sqrt{\boldsymbol{\Sigma}}\propto |s_p|$, cf. Eq.~\ref{eqn:secDervSat}. To avoid too important variations of the sky signal, we only consider, for this specific figure, sky patches which are within the $f_{\rm sky} = 20\%$ at the highest galactic latitudes.}
	\label{fig:full_matrices_delta_beta}
\end{figure}

\begin{table*}[]
	\caption{Specifications of the considered space mission. We assume that angular resolution of each channel is $\sim 1\deg$ i.e. the angular size of the considered frequency maps' pixels, defined with a \textsc{Healpix} $n_{\rm side}^{\rm map}=64$.}
		\begin{center}
		\begin{tabular}{|c|ccccccccccccccc|}
		\hline
		central frequency [GHz] & 40 & 50 & 60 & 68 & 78 & 89 & 100 & 119 & 140 &166 & 195 & 235 & 280 & 337 & 402\\
		\hline
		polarization sensitivity [$\mu$K-arcmin] & 38 & 23 & 15 & 13 & 10 & 9 & 9 & 5 & 5 & 4 & 5 & 5 & 9 & 12 & 21 \\
		\hline
		\end{tabular}
		\end{center}
	\label{table:LB_specs}
\end{table*}

Once the sky patches are defined, in our case they are set by a \textsc{Healpix} grid with $\bar{n}_{\rm side}$, our component separation procedure follows the usual steps, which now have to be performed for each of the patches. Therefore, for each patch, we first solve Eq.~\ref{eq:estimation_of_beta} to estimate the spectral parameters, $\boldsymbol{\tilde{\beta}}$, and then Eq.~\ref{eq:sky_signal_estimate} to recover all the sky components, $\tilde{s}$. The CMB signal recovered in this way, in addition to the actual CMB signal will unavoidably contain some foreground residuals, $\Delta^{\rm CMB}$, as given by Eq.~\ref{eq:stat_fgs_residuals}. If the latter are driven by statistical uncertainties on the estimated spectral indices,  as it is the case when the assumed scaling relations and their sky variability coincide with those of the true sky signals, then the angular power spectrum of foregrounds residuals, $C_\ell^{\rm \boldsymbol{\Sigma}}$, can be expressed as,
\begin{eqnarray}
	\centering
		C_\ell^{\boldsymbol{\Sigma}} \equiv  \sum_{\mathcal{P}}{ \sum_{i,j,k,l}{ \mathbf{\Sigma}^{\mathcal{P}}_{ij}} \boldsymbol{\kappa}_{kl}^{ij,\,\mathcal{P}} C_\ell^{kl,\, \mathcal{P}}},
	\label{eq:Cl_stat_patch}
\end{eqnarray}
which generalizes Eq.~27 of~\cite{Stivoli2010} to the case with multiple sky patches.
Here $\boldsymbol{\kappa}_{kl}^{ij}\equiv \alpha^i_k\alpha^j_l$, and we have assumed no correlation between the estimated $\boldsymbol{\tilde{\beta}}$ and $\tilde{s}$ for any given patch, and no correlation between $\boldsymbol{\delta \beta}$ computed for different patches (see below). The cross spectra of the sky signals, $C_\ell^{kl,\, \mathcal{P}}$ are  noiseless cross-spectra of the actual cosmological and astrophysical sky components, $\tilde{s}$ (Eq.~\ref{eq:s_p_definition}), computed within the sky patch $\mathcal{P}$. $C_\ell^{kl,\, \mathcal{P}}$, like all Fourier quantities involved in this work, are estimated using a simple pseudo-$C_\ell$ computation (using tools provided by the \textsc{Healpy}\footnote{\url{https://github.com/healpy/healpy}} library), and we confirmed our results using \textsc{NaMaster} software\footnote{\url{https://github.com/LSSTDESC/NaMaster}}~\cite{Alonso2018}. Since $E$ and $B$-modes for foregrounds and residuals have roughly similar amplitudes, power spectrum estimation in this particular study does not suffer from e.g. $E$-to-$B$ leakage.
The covariance of errors on spectral indices, $\mathbf{\Sigma}^{\mathcal{P}} \equiv \langle \boldsymbol{\delta \beta}\,\boldsymbol{\delta \beta}^T\rangle$, is estimated using Eq.~\ref{eqn:secDervSat} for each sky patch $\mathcal{P}$.
The sum over patches in Eq.~\ref{eq:Cl_stat_patch} explicitly assumes that statistical residuals are, on average, decorrelated from one sky patch to another, i.e.,
\begin{eqnarray}
	\centering
		&& \left\langle \boldsymbol{\delta\beta}(\mathcal{P}) \right\rangle = \mathbf{0} \label{eq:delta_beta_mean}\label{eq:mean_diagonal_delta_beta}\\
		&& \left\langle \boldsymbol{\delta\beta}(\mathcal{P})  \boldsymbol{\delta\beta}(\mathcal{P}^\prime)^T \right\rangle= \boldsymbol{\Sigma}_{\beta\beta}\ {\rm when\ \mathcal{P} = \mathcal{P}^\prime}\label{eq:diagonal_delta_beta_square}\\
		&& \left\langle \boldsymbol{\delta\beta}(\mathcal{P})  \boldsymbol{\delta\beta}(\mathcal{P}^\prime)^T \right\rangle = \boldsymbol{0} \ {\rm when\ \mathcal{P} \neq\mathcal{P}^\prime}\label{eq:off_diagonal_delta_beta_square}
\end{eqnarray}
where the averages $\left\langle \cdot \right\rangle$ are taken over noise realizations. The decorrelation in Eq.~\ref{eq:off_diagonal_delta_beta_square} is due to the statistical property of the deviations of the recovered $\boldsymbol{\tilde{\beta}}$ from the true values, $\boldsymbol{\beta}^{\rm true}$. This is determined by the foreground signals and the instrumental noise, the latter being assumed to be uncorrelated in this study. 
%\red{discuss about correlated noise !} 
Spatially correlated noise will obviously affect this picture, and break the equality in Eq.~\ref{eq:off_diagonal_delta_beta_square}. Such case is not studied in this work. The updated form of Eq.~\ref{eq:Cl_stat_patch} in this case will have to be generalized, and possibly derived using numerical simulations based on this correlated noise.\\
Eqs~\ref{eq:delta_beta_mean} and~\ref{eq:diagonal_delta_beta_square} are illustrated in Fig.~\ref{fig:histogram_delta_beta}, where we show the distribution of $\boldsymbol{\delta\beta}(\mathcal{P})$ for three spectral parameters, $\boldsymbol{\beta} \in \{ \beta_d, \beta_s, T_d\}$, which we will define in Section~\ref{ssec:simulation_inputs}, in a case where $n^\beta_{\rm side} = \bar{n}^\beta_{\rm side} = 8$. These distributions are centered on zero, and their standard deviations match quite well the diagonal elements of the error matrix, $\boldsymbol{\Sigma}$, averaged over sky patches $\mathcal{P}$.
Eq.~\ref{eq:off_diagonal_delta_beta_square} is illustrated in Fig.~\ref{fig:full_matrices_delta_beta} where we show an estimate of the $\left\langle \boldsymbol{\delta\beta}(\mathcal{P})  \boldsymbol{\delta\beta}(\mathcal{P}') \right\rangle$ matrix, as a function of $\mathcal{P}$ and $\mathcal{P}'$, for $\boldsymbol{\beta}=\beta_d$ (involved in the dust SED as defined in Section~\ref{sssec:input_skies}), averaged over 100 simulations of noise. The diagonal is clearly dominating the entire matrix. 
These observations and their consequences are discussed in detail in section~\ref{sec:results}.\\

\begin{figure*}
	\centering
	\includegraphics[width=0.9\textwidth]{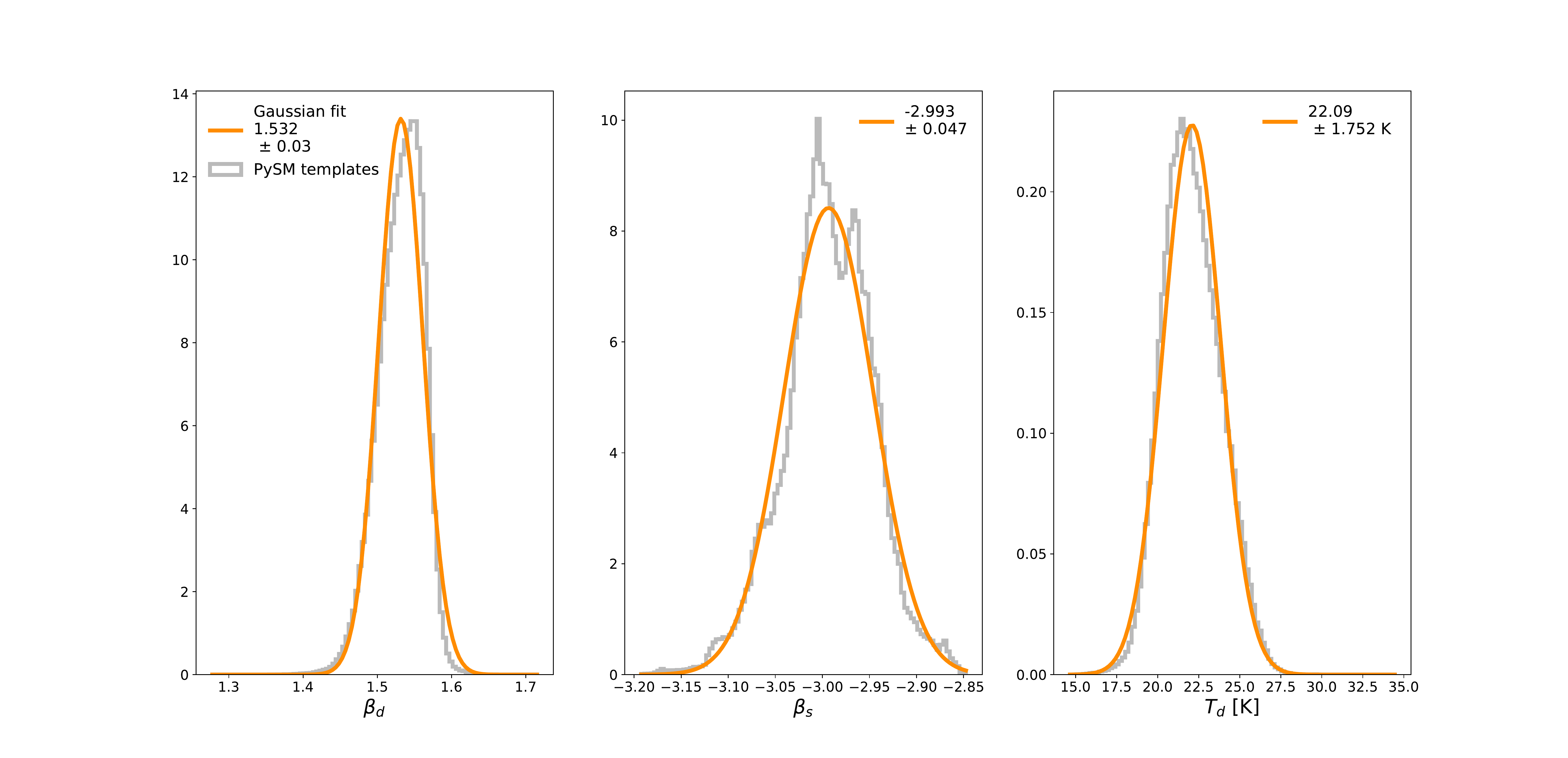}
	\caption{Distributions of spectral indices $\beta_d$, $\beta_s$ and $T_d$ (gray curves) computed from the PySM~\cite{2017MNRAS.469.2821T} ``d1s1'' templates, assuming a $f_{\rm sky}=60\%$ galactic mask. For information, we also show the Gaussian fit of these histograms (orange curves), along with the best-fit means and standard deviations.}
	\label{fig:beta_PySM_distribution}
\end{figure*}

Once we have an estimate of the spectral indices, of the CMB map over the entire sky, $\tilde{s}^{\rm CMB}$, as well as its harmonic coefficients, $\tilde{a}^{\rm CMB}_{\ell\, m}$, the likelihood on tensor-to-scalar ratio $r$ can be expressed as~\cite{Tegmark1997},
\begin{eqnarray}
	\centering
&&\hspace{-0.5cm}-2\log(\mathcal{L}_{\rm cosmo}) = f_{\rm sky}\Big(\sum_{j,j'} (\tilde{a}^{\rm CMB}_j)^\dagger\left[\mathbf{C}^{-1}\right]_{jj'}\tilde{a}^{\rm CMB}_{j'} \Big.\nonumber\\
		&&\hspace{4cm}  \Big. +  \log\left[\det(\mathbf{C})\right] \Big),
	\label{eq:cosmo_likelihood_on_r}
\end{eqnarray}
% \red{effect of fsky above ??}
where $j\equiv \ell^2 + \ell + m$ and $\mathbf{C}$ is the model covariance, which in addition to the standard terms due to the CMB signal and noise, includes the contribution due to the statistical residuals, see Eq.~41 of~\citep{Stompor2016}:
\begin{eqnarray}
	\centering
		\mathbf{C}_{jj'} & \equiv & \langle (\tilde{a}^{\rm CMB}_j)\, (\tilde{a}^{\rm CMB}_j)^\dagger\rangle_{\rm CMB+noise} \nonumber \\
        & = &  \mathbf{C}^{\rm CMB+noise}_\ell \delta_{jj'} + \boldsymbol{\Xi}_{jj'}
	\label{eq:C_ell_def}
\end{eqnarray}
with 
\begin{eqnarray}
	\centering
    	\mathbf{C}^{\rm CMB+noise}_\ell \equiv C_\ell^{\rm prim,\, BB}(r) + C_\ell^{\rm lens,\, BB} + N_\ell
\end{eqnarray}
and where $\boldsymbol{\Xi}_{jj'}$ is the Fourier representation of $\langle s^T\alpha^T\boldsymbol{\Sigma}\alpha s\rangle_{\rm CMB+noise} $, with $s$ the sky signal, $\alpha$ given in Eq.~\ref{eq:alpha_def} and $\boldsymbol{\Sigma}$ in Eq.~\ref{eqn:secDervSat}. As detailed in paragraph~\ref{sssec:input_skies}, the $f_{\rm sky}$ factor in Eq.~\ref{eq:cosmo_likelihood_on_r} is taken to be $60\%$, as we mask the most contaminated regions of the galaxy.
This likelihood can be also derived from the standard likelihood for both the sky component maps and spectral parameters, see e.g., Eq.~5 of~\citep{Stompor2009} via marginalization over the spectral parameters and foreground maps, assuming Gaussian approximation, no correlations between the estimated spectral parameters and sky components, and that the assumed scaling relations for the foregrounds are correct, see Appendix~A of \citep{Stompor2009}. In actual applications, this marginalized covariance would be computed with help of numerical marginalization as the true underlying sky model is not known. For simulations, when the sky model is known, these equations provide a quick way of computing the marginalized likelihood without invoking heavy numerical calculations, a property we capitalize on in this work.

We note that many component separation methods proceed in two steps. First, they estimate the mixing matrix, in a parametric or nonparametric way, which then they employ to compute the estimates of the sky components, Eq.~\ref{eq:sky_signal_estimate}. The recovered CMB map is then used to derive constraints on cosmological parameters often with help of a likelihood function as in Eq.~\ref{eq:cosmo_likelihood_on_r}. This can however only succeed if the covariance assumed in the likelihood includes the statistical residual term. Indeed, as we discuss this later on neglecting it will typically result in significant bias. We find that the statistical residual term can be efficiently computed from the data itself by replacing the true objects by the estimated ones at least in the regime of the experimental sensitivities as considered in this work. \\
While in general, the statistical residual contribution, $\boldsymbol{\Xi}$, to the overall covariance, $\mathbf{C}$, is nondiagonal, it is typically diagonal dominated. Indeed, we find that approximating it as,
\begin{eqnarray}
	\boldsymbol{\Xi}_{jj'} &\approx& C_\ell^{\boldsymbol{\Sigma}} \delta_{jj'},
    \label{eqn:xiDiagApprox}
\end{eqnarray}
affects the final error bar on $r$ on a subpercent level in the specific cases of interest in this paper. For simplicity we will therefore apply this approximation throughout the rest of this work. The relative importance of the off-diagonal terms of $\boldsymbol{\Xi}$ is determined by the magnitude of the error bars on spectral indices, as given by $\boldsymbol{\Sigma}$, with the off-diagonal terms becoming less important with the errors decreasing.  \\
On averaging this likelihood over CMB and noise realizations as in~\citep{Stompor2016} and on defining the ensemble average data covariance, $\mathbf{D}$, as,
\begin{eqnarray}
\mathbf{D}_{j j'} & \equiv &
\left\langle (a^{\rm CMB})^\dagger_{j}\;  a^{\rm CMB}_{j'} \right\rangle_{\rm CMB+noise}
        \approx D_{\ell} \delta_{jj'}.
         \label{eqn:DcovDiagApprox}
\end{eqnarray}
we then obtain, 
\begin{eqnarray}
	\centering
\hspace{-1cm}\big\langle-2\log(\mathcal{L}_{\rm cosmo})\big\rangle_{\rm CMB+noise}  & = & \nonumber \\
&& \hspace{-3.5truecm}= \, f_{sky}\Big({\rm tr}\,\mathbf{C}^{-1}\mathbf{D} +  \log \det\,\mathbf{C}\Big)\ \nonumber \\
&& \hspace{-3.5truecm} \approx f_{\rm sky}\Bigg( \sum_\ell \frac{2\ell+1}{2} \,\mathbf{C}_\ell^{-1}\mathbf{D}_\ell +  \log\left[\det(\mathbf{C})\right] \Bigg).
	\label{eq:cosmo_likelihood_averaged}
\end{eqnarray}
Whenever the assumed covariance, $\mathbf{C}$, provides a good model for the actual data, $\mathbf{D}$, i.e., whenever the assumed and true scaling relations coincide and the statistical residual term is included, the likelihood peaks at the true value of $r$. If a scaling laws mismatch is present, this in general will not be the case, leading to some bias of the best fit value with respect to the true one. However, as long as the bias is smaller than the relevant statistical uncertainty, we will consider the assumed covariance to be sufficient for our purpose. Similarly, neglecting the statistical residuals contribution to the foreground-cleaned CMB map covariance, $\mathbf{C}$, may also result in statistically significant biases, as we discuss this in the following Sections.

Importantly, a computation of the ensemble average likelihood requires only auto- and cross- spectra of some well-defined objects~\citep{Stompor2016}. This is explicitly the case in Eq.~\ref{eq:cosmo_likelihood_averaged}, thanks to the assumptions in Eqs.~\ref{eqn:xiDiagApprox} and~\ref{eqn:DcovDiagApprox}, but it holds more generally when the off-diagonal elements of $\boldsymbol{\Xi}$ and $\mathbf{D}$, are taken into account as shown in~\citep{Stompor2016}. This is in practice very beneficial as these are typically more robustly known than the full covariances of all harmonic modes of these objects, given our limited, present-day knowledge of the foregrounds, and consequently the predictions derived using this formalism are more robust and representative. 

\begin{figure*}
	\centering
	\includegraphics[width=0.8\textwidth]{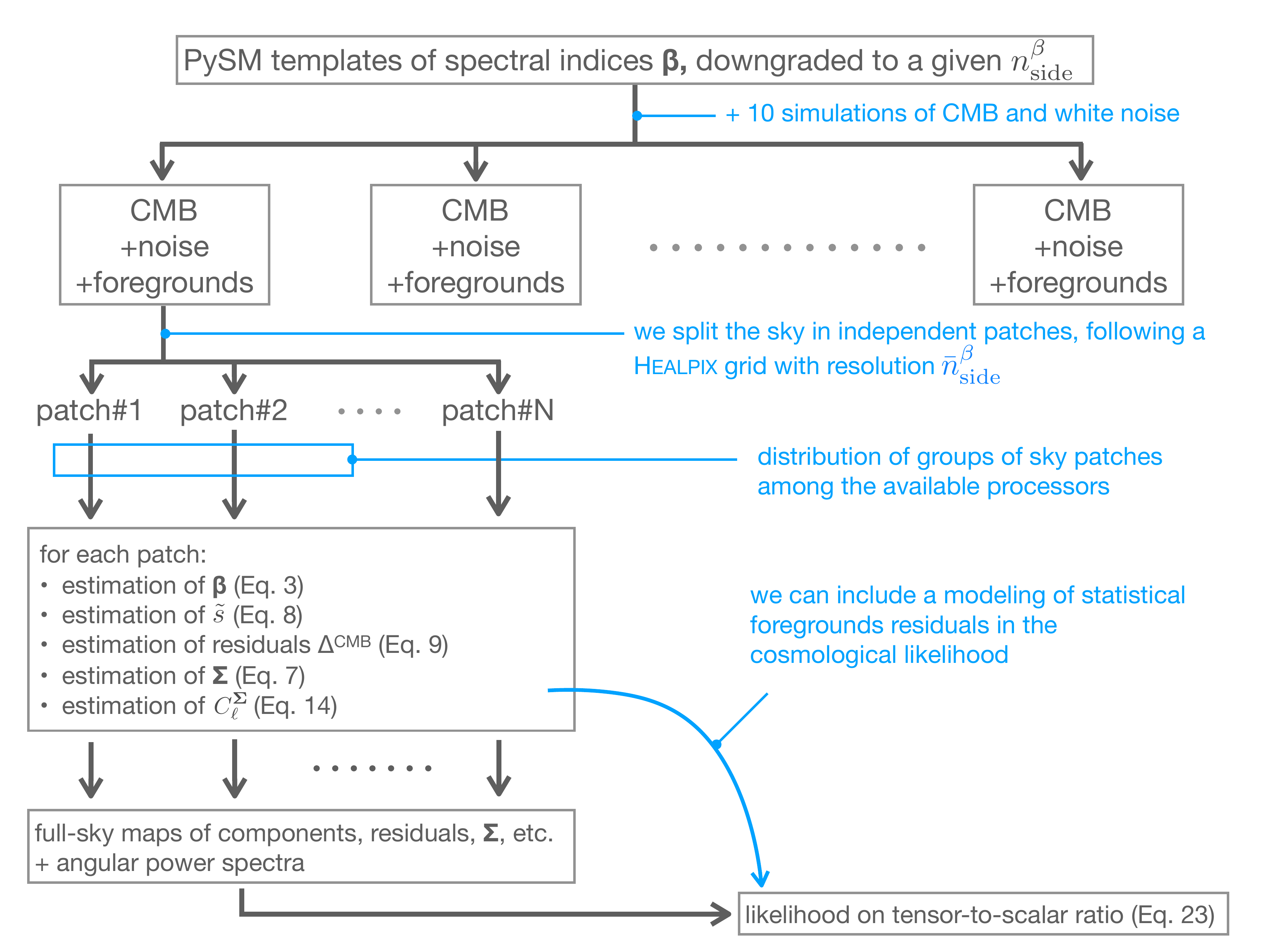}
	\caption{Schematic description of our method and its implementation.}
	\label{fig:explanatory_schematic}
\end{figure*}

\subsection{Simulations setup}
\label{ssec:simulation_inputs}

\subsubsection{Instrument}

We consider a full-sky space mission, equipped with 15 frequency bands between 40 and 400GHz, following the specifications summarized in Table~\ref{table:LB_specs}. These roughly correspond to a typical large scale satellite mission as considered for a deployment within the next decade.\\
Without loss of generalities, we consider $\delta$ bandpasses in this work --- although the presented formalism is transparent to more realistic bandpasses. That said, one should keep in mind that imperfect bandpasses can significantly affect the performance of the parametric component separation~\citep{2018arXiv180307630W,2017JCAP...12..015T,Errard2012}. 

\subsubsection{Input skies}
\label{sssec:input_skies}

We use the templates of spectral parameters $\boldsymbol{\beta}$ from the ``d1s1'' model of PySM~\cite{2017MNRAS.469.2821T}, which are degraded and smoothed down to angular scales corresponding to a \textsc{Healpix} pixelization ranging from $0\leq n_{\rm side}^\beta \leq 64$. This $n_{\rm side}^\beta$ defines the ``true'' sky patches, in opposition to the ones used in the component separation, defined by $\bar{n}_{\rm side}^\beta$. Fig.~\ref{fig:beta_PySM_distribution} shows the distribution of spectral indices, $\beta_d$, $\beta_s$ and $T_d$ as obtained with a $f_{\rm sky}=60\%$ sky mask. Given these templates, we also use the PySM maps of dust and synchrotron to generate sets of noisy frequency $\{Q, U\}$ maps following Eq.~\ref{eq:data_modeling}. 
Note that although we degrade and smooth the $\boldsymbol{\beta}$ maps to $2\leq n_{\rm side}^\beta \leq 64$, we consistently keep the final resolution of the frequency maps, $d_p^{{\rm obs,}\,\nu}$, set to $n^{\rm map}_{\rm side}$ = 64. 
Throughout this work, we apply a $60\%$ galactic mask, as constructed for the analysis of the Planck's HFI dust maps.\\
Except when mentioned otherwise, we assume a power law for synchrotron emission and a modified black body for dust, expressed as:
\begin{eqnarray}
	\centering
		\mathbf{A}^{\rm sync}(\nu,\beta_s) &\equiv& \nu^{\beta_s}\label{eq:sync_SED}\\
		\mathbf{A}^{\rm dust}(\nu,\beta_d,T_d) &\equiv& \left(\frac{h\nu}{kT_d}\right)^{\beta_d}B_\nu(T_d)\label{eq:dust_SED}
\end{eqnarray}
where $B_{\nu}$ is a black body at a temperature $T_d$, and all spectral indices $\boldsymbol{\beta}\in\{\beta_d,\beta_s,T_d\}$ are dependent on the considered sky patch, $\boldsymbol{\beta} = \boldsymbol{\beta}(\mathcal{P})$.\\

\subsubsection{Implementation}
The implementation of the method detailed in section~\ref{ssec:generalization_multipatch} is illustrated in Fig.~\ref{fig:explanatory_schematic}. The process can be summarized as follows:
\begin{itemize}
	\item we generate foregrounds-only frequency maps using the PySM templates for dust, synchrotron, $\beta_d$, $\beta_s$ and $T_d$. We degrade the dust and synchrotron templates to $n^{\rm map}_{\rm side} = 64$, and the $\boldsymbol{\beta}$ templates to  $2\leq n^\beta_{\rm side}\leq n^{\rm map}_{\rm side}$ --- each case is then independently analyzed and described in section~\ref{sec:results};
	\item for each foregrounds simulation, we generate 10 simulations of CMB (assuming Planck's best-fit  fiducial $\Lambda$CDM parameters~\cite{2016A&A...594A..13P}, with\footnote{Changing the reionization optical depth $\tau$ from $0.055$ to $0.050$ changes the final uncertainty on tensor-to-scalar ratio $r$ by $\leq 10\%$.} $\tau=0.055$ and $r=0.001$) and white instrumental noise (following the specification detailed in Table~\ref{table:LB_specs});
	\item we then distribute, to the available CPUs, groups of sky patches which are given by each \textsc{Healpix} pixel with a $\bar{n}^\beta_{\rm side}$ resolution (equal or different from $n^\beta_{\rm side}$); 
	\item for each patch and each CMB+noise simulation, we estimate the spectral indices, the associated error matrix and the sky signals (CMB, dust and synchrotron $\{Q, U\}$ maps);
	\item once all patches and all simulations have been analyzed, we combine all sky patches to produce a nearly full sky map of reconstructed sky signals (CMB, dust, synchrotron), as well as noise, spectral indices, $\boldsymbol{\Sigma}$, etc.;
	\item we then estimate the angular power spectra $\tilde{C}_\ell$ of these full-sky components, residuals, noise, etc., avoiding the low galactic regions --- as mentioned earlier, we apply Planck's $f_{\rm sky} = 60\%$ galactic mask;
    \item we compute $C_\ell^{\boldsymbol{\Sigma}}$, Eq.~\ref{eq:Cl_stat_patch}, using $\boldsymbol{\Sigma}$ and the angular power spectra estimated at the previous steps;
	\item we finally evaluate and optimize the likelihood on tensor-to-scalar ratio $r$, Eq.~\ref{eq:cosmo_likelihood_averaged}, and estimate the corresponding $68\%$ C.L. error bar. When estimating $\mathcal{L}(r)$, we perform two runs, including the $C_\ell^{\boldsymbol{\Sigma}}$ term or not in the modeled covariance, cf. the $\boldsymbol{\Xi}$ term in Eq.~\ref{eq:C_ell_def}.
\end{itemize}

%%%%%%%%%%%%%%%%%%%%%%%%%%%%%%%%%%%%
\section{Results}
\label{sec:results}

We present in this section the results of running the previously detailed analysis for several values of $n^\beta_{\rm side}$ (defining the true $\boldsymbol{\beta}$) and $\bar{n}^\beta_{\rm side}$ (defining the patches for the analysis):
\begin{itemize}
	\item first, in Section~\ref{ssec:ideal_case}, in the case of a ``simple'' and ideal sky i.e. when patches used during the component separation, defined with $\bar{n}^\beta_{\rm side}$, match the ones used to simulate the sky, $n^\beta_{\rm side}$; 
	\item second, in Section~\ref{ssec:realistic_case}, in the case of a systematic mismatch either between the SED parametrization or between the analysis patch and the true spatial variations of spectral indices considered in the simulation i.e. $\bar{n}^\beta_{\rm side} \neq n^\beta_{\rm side}$. 
\end{itemize}
These two cases allow us to probe two regimes for the foregrounds residuals, $\Delta^{\rm CMB}$, Eq~\ref{eq:resBreakDown}, statistical and systematic, and the potential trade-offs between the two.
From this perspective, the focus of Section~\ref{ssec:ideal_case} is on the statistical residuals, $\Delta_{\rm stat}^{\rm CMB}$, whereas the systematic residuals, $ \Delta_{\rm sys}^{\rm CMB}$, are considered in Section~\ref{ssec:realistic_case}.

\subsection{Ideal cases for which $n_{\rm side}^\beta=\bar{n}_{\rm side}^\beta$}
\label{ssec:ideal_case}

\begin{figure}
	\centering
	\includegraphics[width=1.0\columnwidth]{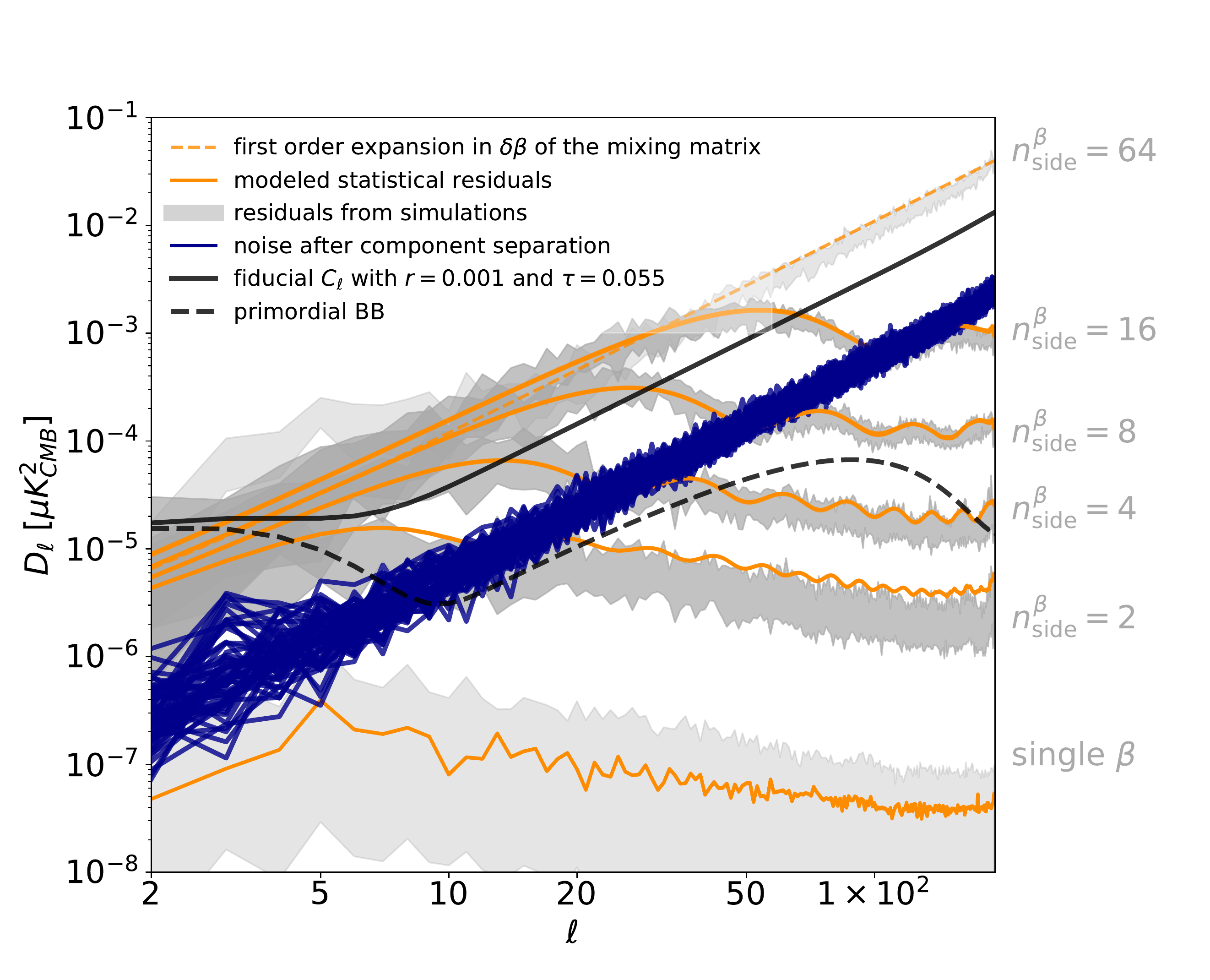}
	\caption{Angular power spectra of the noiseless foreground residuals, gray curves, Eq.~\ref{eq:tot_fgs_residuals}, and the noise after component separation, blue solid lines, in the ideal case where the patches used in the analysis match the input patches, i.e. $n_{\rm side}^\beta = \bar{n}_{\rm side}^\beta$, and obtained from 10 different CMB+noise simulations for each value of $n_{\rm side}^\beta$. The modeled statistical curves, in orange, are computed using Eq.~\ref{eq:Cl_stat_patch}.   The ``single $\beta$'' case, orange dashed line, corresponds to the case when a single set of spectral parameters, $\boldsymbol{\beta}$ is fit for and the spatial variability of the spectral indices is modeled as an extra sky component, resulting from the first order expansion of the foreground scaling laws around the mean values of the parameters as in Eq.~\ref{eq:stolyarov}, see, e.g.~\citet{Errard2011}.
    Finally, the black solid and dash curves are the theoretical total and primordial $B$-modes, assuming $r=0.001$ and $\tau=0.055$.
    }
	\label{fig:power_spectra_figure}
\end{figure}

In the ideal case where the sky patches used in the analysis match the ones assumed to simulate the sky, the performance of the component separation is expected to be limited by the statistical errors associated with the estimation of spectral indices.  
This can be the case when the modeling has enough degrees of freedom, and when the instrument has enough frequency coverage and sensitivity to break degeneracies between spectral indices. 
The fact that $n_{\rm side}^\beta=\bar{n}_{\rm side}^\beta$ put us in the situation where the spectral analysis has exactly the right number of degrees of freedom: in our study, three free parameters, $\beta_d$, $\beta_s$ and $T_d$, for each sky patch. In addition, as detailed in~\citet{Errard2011}, $\boldsymbol{\Sigma}$ is primarily driven by the distribution of sensitivities and frequencies of the instrument.

In the case of $n_{\rm side}^\beta=\bar{n}_{\rm side}^\beta=8$, Fig.~\ref{fig:histogram_delta_beta} shows the distribution of the deviations between the recovered $\boldsymbol{\beta}$ and the true ones. These distributions are centered on zero with a standard deviation well approximated by $\boldsymbol{\Sigma}$. Using the same type of sky simulations, Fig.~\ref{fig:full_matrices_delta_beta} shows the $\langle \delta\beta_d(\mathcal{P}) \delta\beta_d(\mathcal{P}^\prime)\rangle$ matrix, averaged over 100 simulations of CMB and noise.

Fig.~\ref{fig:power_spectra_figure} shows the power spectrum of foregrounds residuals, Eq.~\ref{eq:tot_fgs_residuals}, after performing the component separation independently in patches following a \textsc{healpix} pixelization with $ 2 \leq n^\beta_{\rm side}=\bar{n}^\beta_{\rm side} \leq 64$. As already mentioned, under these specific assumptions, residuals are purely sourced by statistical errors on the recovered $\boldsymbol{\beta}$. The properties of these residuals are driven by the properties of $\boldsymbol{\delta\beta}$, Eqs.~\ref{eq:mean_diagonal_delta_beta},~\ref{eq:diagonal_delta_beta_square} and~\ref{eq:off_diagonal_delta_beta_square}, and by the amplitudes and spatial distributions of foregrounds signals. We also show the noise power spectrum after component separation, and the modeled statistical foregrounds residuals, $C_\ell^{\boldsymbol{\Sigma}}$, estimated from $\boldsymbol{\Sigma}$ and $\tilde{C}_\ell$ in each sky patch. 

\begin{figure}
	\centering
		\includegraphics[width=\columnwidth]{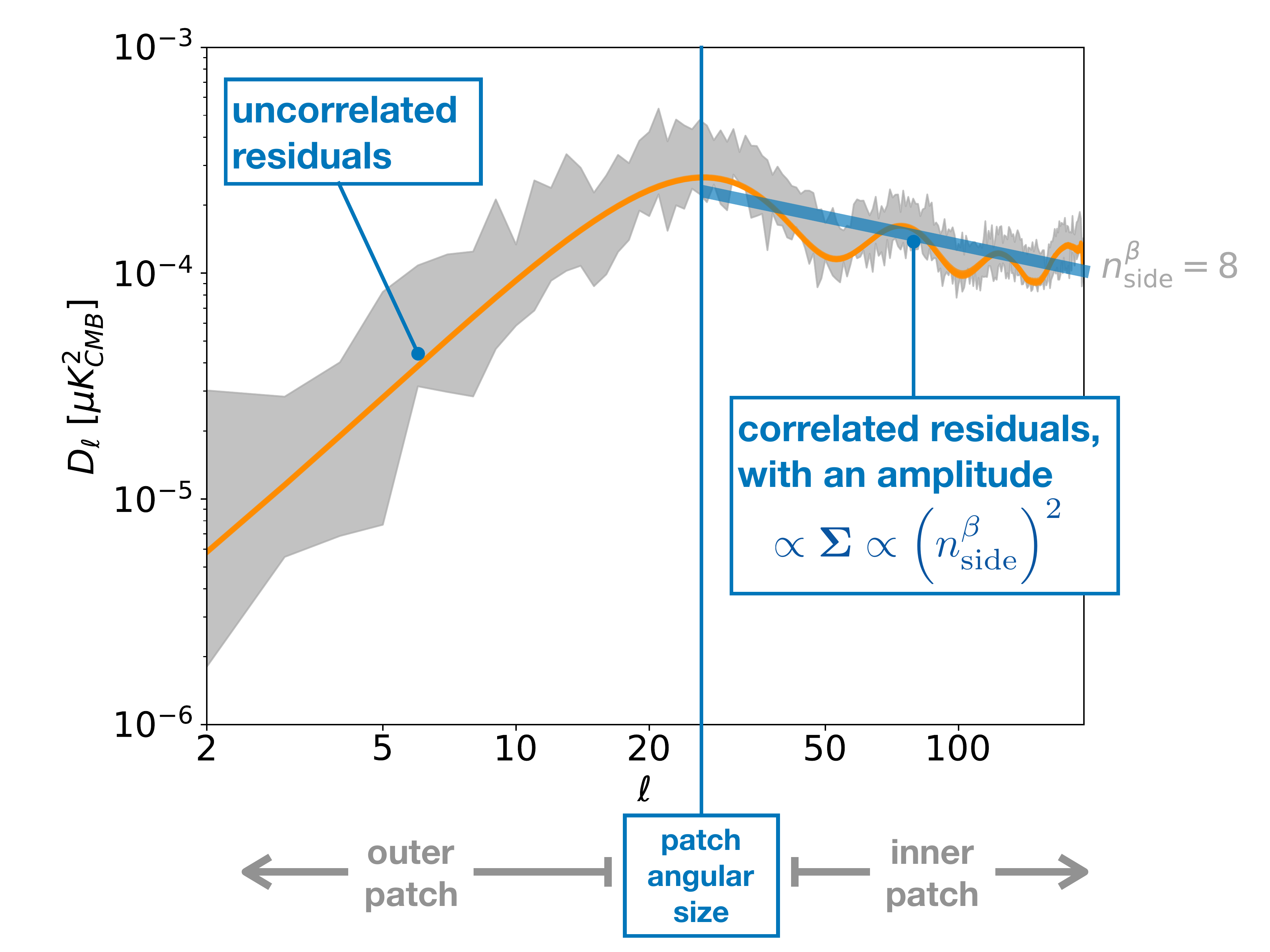}
	\caption{Schematic explanation of the angular power spectrum of statistical foregrounds residuals. The largest angular scales, above the patch size (here $n_{\rm side}^\beta = \bar{n}_{\rm side}^\beta = 8$) , are uncorrelated and follow a white noise spectrum, $\propto \ell^2$. This is true until it reaches scales which are comparable or smaller than the patch size ($\ell_{\rm patch} \approx 25$ for $n_{\rm side}^\beta = 8$): the angular power spectrum of the residuals then becomes a red spectrum, proportional to the input foregrounds power spectra.}
	\label{fig:expanatory_figure}
\end{figure}

Fig.~\ref{fig:expanatory_figure} picks one residuals power spectrum and one model curve from Fig.~\ref{fig:power_spectra_figure} and provides an interpretation for the different parts of the spectrum. On the largest angular scales, as it was anticipated by Eq.~\ref{eq:off_diagonal_delta_beta_square} and Fig.~\ref{fig:full_matrices_delta_beta}, we observe a decorrelation of the residuals. By its definition, this is captured by the orange curve, which follows the expression of the semi-analytical statistical foregrounds residuals, $C_\ell^{\boldsymbol{\Sigma}}$, Eq.~\ref{eq:Cl_stat_patch}.
Once the patch angular size is reached ($\ell_{\rm patch} \sim 25$ for $n^\beta_{\rm side}=\bar{n}^\beta_{\rm side}=8$), there is a change of slope in the shape of the foregrounds residuals, which are transitioning from the outer ($\ell\leq \ell_{\rm patch}$) to the inner ($\ell\geq \ell_{\rm patch}$) patch angular scales, where residuals become basically proportional to the input foregrounds power spectra i.e. typically $C_\ell \propto \ell^{-2.4-2.6}$~\cite{Planck_XXX_2016}. On these small angular scales, up to some factors depending on $\tilde{\mathbf{A}}$ and its derivatives, the coefficient of proportionality is given by the error bars on spectral indices, $\boldsymbol{\Sigma}$, as detailed in Eq.~\ref{eqn:secDervSat}. The oscillatory behavior of the residuals over the inner-patch size correspond to harmonics of the chosen patch size. This is a result of the estimated values of the spectral parameters, $\boldsymbol{\beta}$, abruptly changing between the patches. In more realistic circumstances, where the sizes of different patches would typically vary, these oscillations would be smoothed out.\\
The modeled statistical foregrounds residuals (orange curves) in Fig.~\ref{fig:power_spectra_figure} are  consistent with the observed residuals estimated from simulations (light gray curves), up to the cases for which $n_{\rm side}^\beta \leq 16$. Beyond that value, the signal-to-noise per sky patch becomes too poor to estimate $\boldsymbol{\beta}$ parameters given the instrumental specifications, Table~\ref{table:LB_specs}, and leads to slightly biased estimates of $\boldsymbol{\beta}$, breaking the relations in Eqs.~\ref{eq:mean_diagonal_delta_beta},~\ref{eq:diagonal_delta_beta_square} and~\ref{eq:off_diagonal_delta_beta_square}. Although the assumed modelling of the SEDs is correct, poor signal-to-noise leads to numerical instabilities of the spectral likelihood optimization. Including priors on $\boldsymbol{\beta}$ could certainly mitigate some of these cases. This can be particularly seen in Fig.~\ref{fig:power_spectra_figure} with the light gray curves for the $n_{\rm side}^\beta = 64$ case, which shows a slight excess of power on the largest angular scales. 
This situation could be improved with help of theoretical or observational priors imposed on the spectral parameters $\boldsymbol{\beta}$, what is however not explored in this article.
The angular power spectra of the noise after component separation are unchanged from case to case, which is expected in this formalism: noise levels are given by the CMB element of the noise covariance $\left(\tilde{\mathbf{A}}^T\mathbf{N}^{-1}\tilde{\mathbf{A}} \right)^{-1}$, cf. Eq.~\ref{eq:sky_signal_estimate}, and this amplitude does not depend on the chosen sky patches.\\
Finally, for completeness, we show in Fig.~\ref{fig:power_spectra_figure}  that the residuals obtained in the multipatch approach in the case $\bar{n}_{\rm side}^\beta = n_{\rm side}^\beta = 64$ correspond to the noise after component separation, when the spatial variability of the foreground properties is accounted for by expanding the mixing matrix to first order in $\boldsymbol{\delta \beta}$, i.e.,
\begin{eqnarray}
	\centering	
    	\mathbf{A}(\boldsymbol{\beta}) \approx \mathbf{A}(\langle\boldsymbol{\beta}\rangle) + \boldsymbol{\delta \beta}\left.\frac{\partial \mathbf{A}}{\partial \boldsymbol{\beta}}\right|_{\langle\boldsymbol{\beta}\rangle},
    \label{eq:stolyarov}
\end{eqnarray}
and recasting the data model in Eq.~\ref{eq:data_modeling} by introducing an extra sky component, $\boldsymbol{\delta \beta}\, \mathbf{s}_p$, and an extended mixing matrix $\mathbf{A'} \equiv [\mathbf{A}{(\langle \boldsymbol{\beta}\rangle}),\; \partial \mathbf{A}/\partial \boldsymbol{\beta}|_{\langle\boldsymbol{\beta}\rangle}]$. The extra sky component results in a higher noise in the component-separated CMB map, i.e. the (CMB, CMB) elements of the $({\mathbf{A'}}^T\mathbf{N}^{-1}\mathbf{A'})^{-1}$ matrix. This boosted noise is illustrated as the dashed, orange curve in Fig.~\ref{fig:power_spectra_figure} and matches very well the foregrounds residuals obtained when performing our analysis over $\bar{n}_{\rm side}^\beta = 64$ independent sky patches showing that both these approaches lead to similar statistical uncertainty of the recovered CMB map but only if both sources of statistical errors, noise and statistical residuals, are taken into account.

\begin{figure*}
	\centering
		\includegraphics[width=\textwidth]{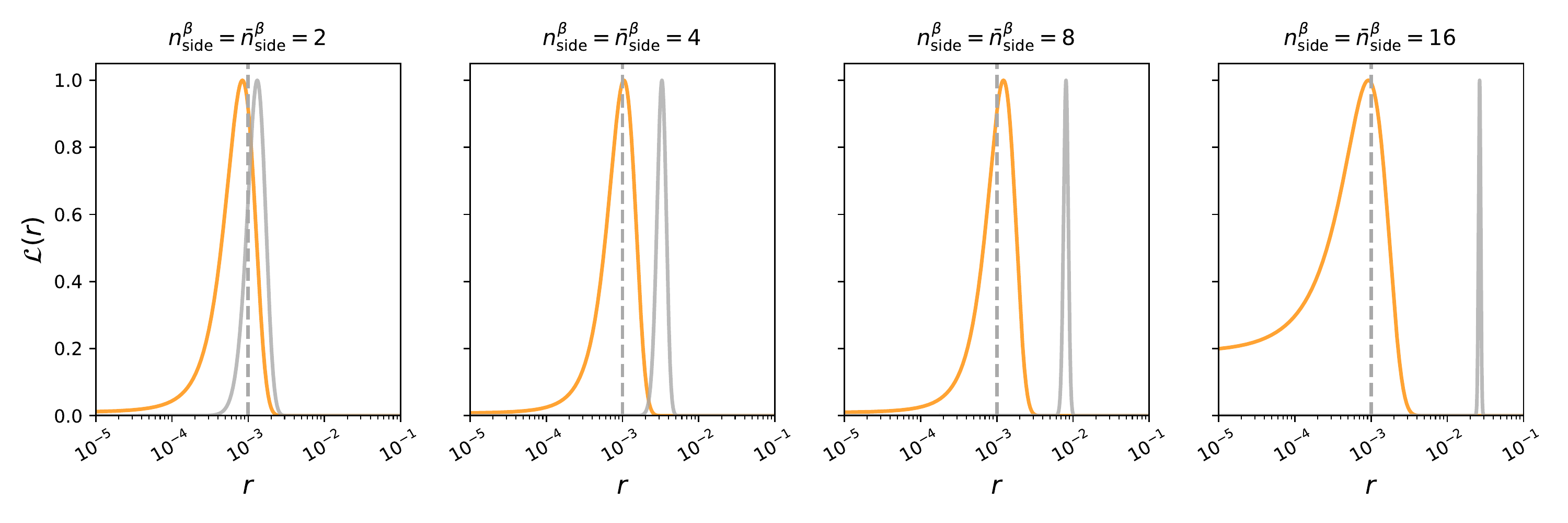}
	\caption{Likelihoods on tensor-to-scalar ratio, $r$, following Eq.~\ref{eq:cosmo_likelihood_averaged}. The gray curves are obtained while neglecting the statistical residual term, i.e., $C_\ell^{\boldsymbol{\Sigma}}$, in Eq.~\ref{eq:C_ell_def}, while the orange curves are derived when the modeled statistical foregrounds residuals are taken into account. The true value of $r=0.001$, used in the input simulations, is represented as the vertical gray dash lines.}
	\label{fig:likelihoods_ideal_case}
\end{figure*}

The post component separation products are then propagated to the cosmological likelihood, Eq.~\ref{eq:cosmo_likelihood_averaged}. The likelihoods are computed over a range of tensor-to-scalar ratios, as shown in Fig.~\ref{fig:likelihoods_ideal_case}. Each panel of the  figure corresponds to a different value of $\bar{n}_{\rm side}^\beta = n_{\rm side}^\beta$. The gray curves correspond to the cases where we do not include the statistical residual term in the modeled total covariance $\mathbf{C}$ of the CMB signals, Eq.~\ref{eq:C_ell_def}. The orange curves are computed with this term included. As it could be expected from the amplitudes of the foregrounds residuals at the power spectrum level in Fig.~\ref{fig:power_spectra_figure}, the effect of including $C_\ell^{\boldsymbol{\Sigma}}$ is crucial to ensure no bias in the recovered tensor-to-scalar ratio~\cite{2017MNRAS.468.4408H}, which is taken as $r=0.001$ in the input sky simulations. The bias on $r$, seen in the cases with the statistical residuals term neglected, follows the amplitude of statistical foregrounds residuals: the higher $\bar{n}_{\rm side}^\beta = n_{\rm side}^\beta$, the larger the residuals are and the larger is the bias on $r$. For $\bar{n}_{\rm side}^\beta = n_{\rm side}^\beta = 8$, we see that we recover $r\approx 0.008$ (instead of the input $r=0.001$) with $\sigma(r) \sim 5\times 10^{-4}$ in both these cases. More generally, the uncertainty on $r$, $\sigma(r)$, vary from $\approx 4\times 10^{-4}$ for $\bar{n}_{\rm side}^\beta = n_{\rm side}^\beta = 2$ up to $ \approx 7\times 10^{-4}$ for $\bar{n}_{\rm side}^\beta = n_{\rm side}^\beta = 16$. This slight increase  of the constraints on $r$ is due to the increase of the final variance, due to the power of statistical foregrounds residuals, which scale as $\boldsymbol{\Sigma} \propto \left( {\rm \#\,of\,pixels\,per\,patch}\right)^{-1} \propto ( \bar{n}_{\rm side}^\beta)^{2}$, cf. Fig~\ref{fig:expanatory_figure}.

\subsection{Cases for which $n_{\rm side}^\beta\neq\bar{n}_{\rm side}^\beta$}
\label{ssec:realistic_case}

We now explore cases for which the assumed parametrization of the SEDs is systematically incorrect, either because the true scaling relations are inherently more complex, or because of the unaccounted for spatial variability of the foreground scaling, $\bar{n}_{\rm side}^\beta \neq n_{\rm side}^\beta$.\\
In such cases, the foreground residuals are not only sourced by  statistical errors on $\boldsymbol{\beta}$, as described in the previous section, but also by systematic errors. The mismatch between the true and assumed foreground properties will typically break the equalities in Eqs.~\ref{eq:mean_diagonal_delta_beta},~\ref{eq:diagonal_delta_beta_square} and ~\ref{eq:off_diagonal_delta_beta_square}, and in particular lead to $\langle\boldsymbol{\delta\beta}\rangle \neq \mathbf{0}$.\\

To illustrate these systematics effects, we produce two new sets of foregrounds simulations:
\begin{itemize}
	\item \textbf{sim\#1 --- } frequency maps with $\boldsymbol{\beta}\,\in\,\{\beta_d, T_d,\beta_s\}$ are varying on a $n_{\rm side}^\beta = 64$ \textsc{Healpix} grid, but we keep $2\leq \bar{n}_{\rm side}^\beta \leq 8$ in the component separation method. This way, the effective SEDs, averaged over the analysis patches, will not follow Eqs.~\ref{eq:sync_SED} and~\ref{eq:dust_SED} anymore, but would introduce effective curvatures~\cite{2017MNRAS.472.1195C};
	\item \textbf{sim\#2 --- } frequency maps using the option ``a2d6s3'' of PySM --- corresponding to 2\% polarized AME, dust model from~\citet{Vansyngel2017} and curved synchrotron following \citet{2012ApJ...753..110K}, with all the frequency scaling laws spatially varying over a $n_{\rm side}^\beta = 64$ \textsc{Healpix} grid. We assume $\bar{n}_{\rm side}^\beta = 4$ in the component separation method.
\end{itemize}

\begin{figure*}
	\centering
		\includegraphics[width=2\columnwidth]{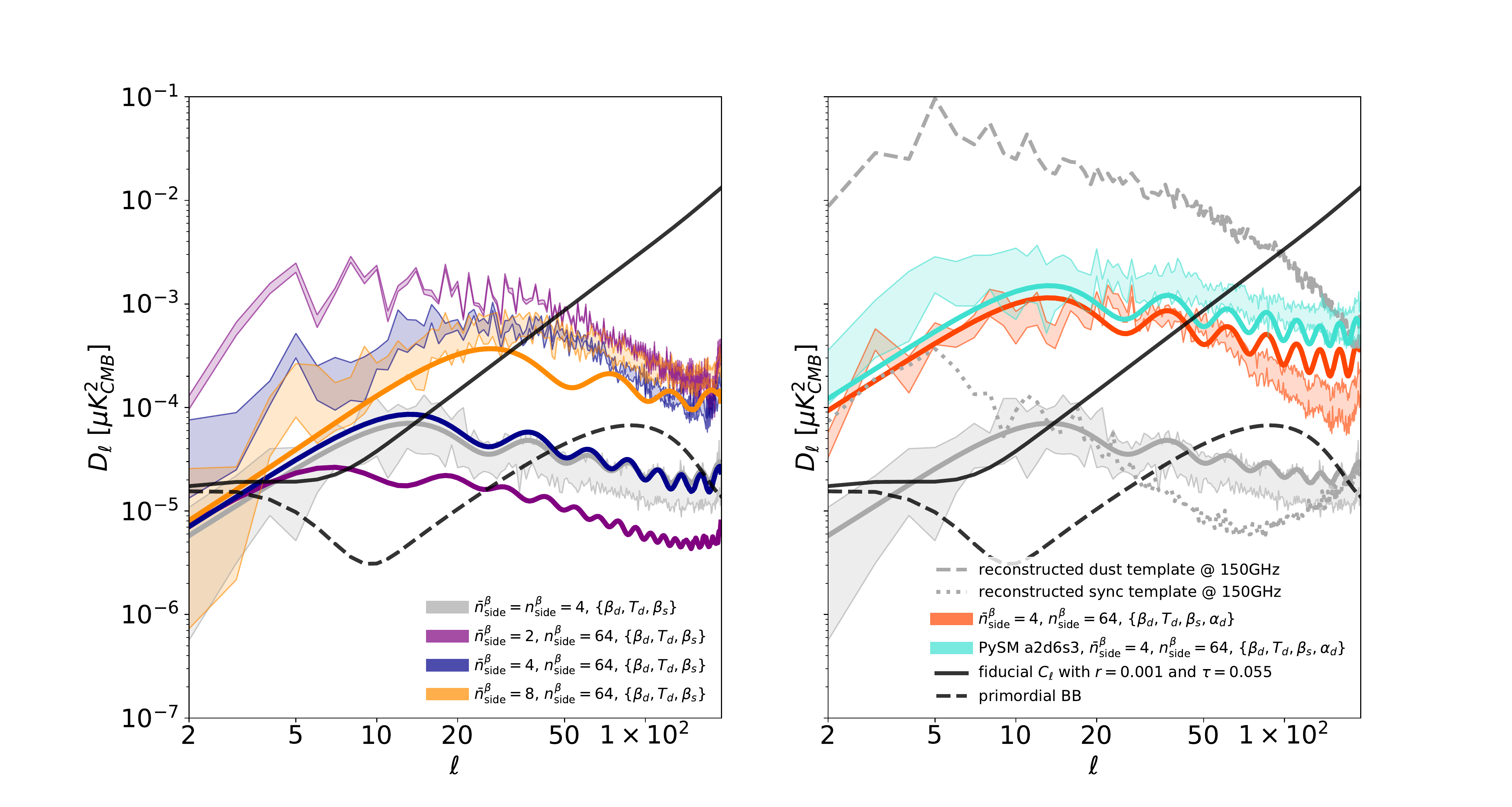}
		\caption{As Fig.~\ref{fig:power_spectra_figure} but for complex foregrounds simulations, described in section~\ref{ssec:realistic_case}. Colored bands show the scatter of foreground residuals obtained for a set of 10 independent CMB+noise simulations implementing different foreground models and/or component separation assumptions as described in the legends. 
        The results shown in the left panel are for \textbf{sim\#1} and thus assume that $n^\beta_{\rm side} > \bar{n}^\beta_{\rm side}$. The right panel depicts \textbf{sim\#2}, i.e., when $n^\beta_{\rm side}= 64$ and $\bar{n}^\beta_{\rm side}=4$, and the input foregrounds include more complex emissions than previously assumed.
        The color solid lines show the analytic predictions for the statistical residuals computed for each of the considered cases.
        The gray bands are added for reference in both panels and depict the $\bar{n}^\beta_{\rm side} = n^\beta_{\rm side} = 4$ curves shown in Fig.~\ref{fig:power_spectra_figure}. The dashed and dotted gray lines in the right panel show the reconstructed dust and synchrotron templates corresponding to $\tilde{C}_\ell^{\rm (d\times d)}$ and $\tilde{C}_\ell^{\rm (s\times s)}$, i.e., the auto power spectrum of the reconstructed, nearly full sky dust and synchrotron maps, as appearing in Eq.~\ref{eq:modeling_of_Cl_sys}.}
	\label{fig:power_spectra_figure_complex_sky}
\end{figure*}

We show in Fig.~\ref{fig:power_spectra_figure_complex_sky} the foregrounds residuals obtained after running the component separation on the two foregrounds simulations described above. For reference, we also show the case where $\bar{n}^\beta_{\rm side} = n^\beta_{\rm side} = 4$, referred to as fiducial in the following and the one already depicted in Fig.~\ref{fig:power_spectra_figure}. These two new sets of foregrounds simulations allow us to probe two distinct systematic effects that are described below.\\

\begin{figure*}
	\centering
		\includegraphics[width=0.8\textwidth]{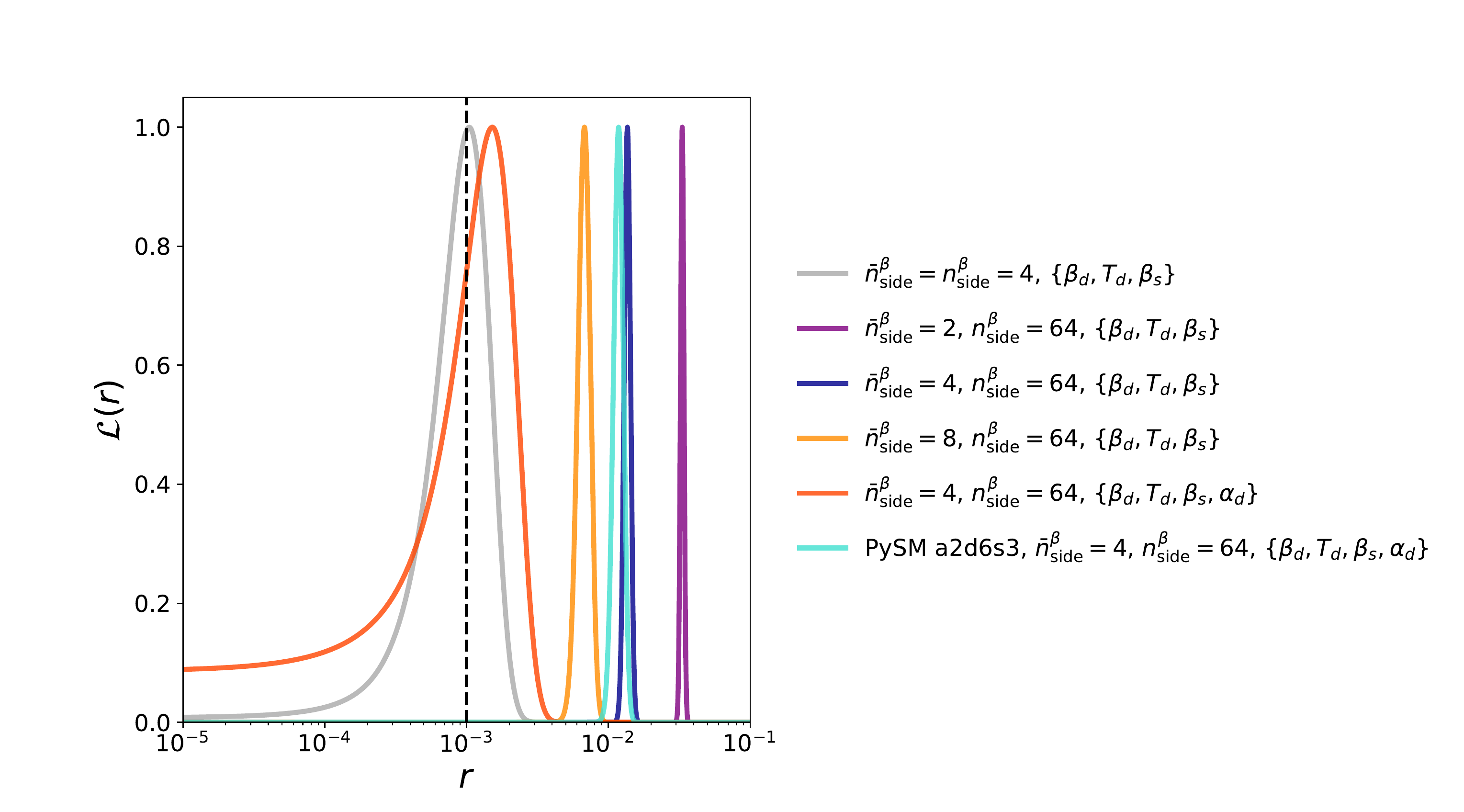}
	\caption{Likelihood functions for the tensor-to-scalar ratio, r, $\mathcal{L}_{\rm cosmo}(r)$, computed for different foreground models incorporating sub-degree variations of the spectral indices on \textsc{HEALpix} pixels with $n_{\rm side}^\beta = 64$ and for different component separation assumptions. The purple, blue and orange curves are the likelihoods when the analysis patches have a size $\bar{n}_{\rm side}^\beta = 2$, $4$ and $8$ respectively. The deep orange curve corresponds to the $\bar{n}_{\rm side}^\beta = 4$ case with an extra free running index assumed to model the dust SED in the component separation modeling. Finally, the turquoise curve is obtained when running this exact same component separation on a more complex sky, the ``a2d6s3'' PySM simulations as described in Sec.~\ref{ssec:realistic_case}.}
	\label{fig:likelihood_complex_sky_sdrun}
\end{figure*}

In \textbf{sim\#1}, spectral indices $\boldsymbol{\beta}$ are spatially varying on a denser \textsc{healpix} grid than the analysis one, $n_{\rm side}^\beta = 64 > \bar{n}_{\rm side}^\beta$. 
The resulting foreground residuals computed for these cases are displayed as color bands in the left panel of Fig.~\ref{fig:power_spectra_figure_complex_sky} and show significantly more power not only with respect to the fiducial case, but also as compared to the statistical residuals which are calculated analytically as before and depicted with color, solid lines. We note that while the statistical residuals decrease with the decreasing value of $\bar{n}_{\rm side}^\beta$ and thus an increasing size of the patches assumed for the recovery, the total residuals as calculated numerically from the simulations tend to increase in particular at the low-$\ell$ part of the spectra. This is expected as the larger patches lead to smaller statistical uncertainties on the recovered spectral parameters, i.e., smaller statistical residuals, but also to larger discrepancies between the assumed and effective foreground scaling relations, i.e., larger systematic residuals, and these are the latter which quickly dominate. 
Fig.~\ref{fig:likelihood_complex_sky_sdrun} shows the corresponding likelihood obtained for the various cases $\{n^\beta_{\rm side},\bar{n}^\beta_{\rm side}\} \in \left\{\{4,4\},\{64,2\},\{64,4\},\{64,8\}\right\}$. Unsurprisingly, even when including the modeled statistical foregrounds as we detailed in Section~\ref{ssec:ideal_case} and Fig.~\ref{fig:likelihoods_ideal_case}, the bias on tensor-to-scalar ratio $r$ is very significant, driven by the unaccounted for in any way systematic residuals. 
One possibility to mitigate these effects at the component separation level is to include extra free spectral parameters in the effective modeling of dust and synchrotron SEDs, as they, due to averaging of different SED within the patches, no longer correspond to simple power laws or modified black bodies.
We explore here an idea developed in~\citet{2017MNRAS.472.1195C}, which proposes an expansion with moments of the SEDs to model the effect of spatially averaging nonlinear functions, like power law or modified black body. They analytically demonstrated that this average over different line-of-sight leads to a modification of the scaling laws --- typically, for $\mathbf{A}$ being a power-law,
\begin{eqnarray}
	\centering
	        \langle \mathbf{A}^\nu \rangle_{\rm sky\, pixels} \approx \nu^{\beta+ \alpha \log\left(\nu\right) + \alpha^\prime \left[\log\left(\nu\right)\right]^2 + ... }.
	\label{eq:moments_expansion}
\end{eqnarray}
In our study, driven by the limited signal-to-noise per sky patch, we decide to expand the list of free $\boldsymbol{\beta}$ by only one new parameter --- a running index, $\alpha_d$, in the dust emission law. This parameter, as the other $\{\beta_d,\beta_s,T_d\}$ , is to be adjusted in each sky patch, assuming here $\bar{n}_{\rm side}^\beta = 4$ for the analysis patches and $n_{\rm side}^\beta = 64$ to produce the sky simulations. 
One should note that adding too many free spectral parameters could affect the Gaussian behavior in Eq.~\ref{eq:diagonal_delta_beta_square}, i.e. the fact that error bars on spectral parameters can be simply approximated by the curvature of the spectral likelihood. This could lead to a wrong estimation of the statistical foregrounds residuals, and sometimes leading to significant degeneracies affecting the optimization of $\mathcal{L}_{\rm spec}$, Eq.~\ref{eq:spectral_likelihood}.
When including the free running index for the dust SED, we also include a prior $\Pi$ on $\alpha_d$, in order to keep the posterior approximately Gaussian,
\begin{eqnarray}
	\Pi_{\alpha_d} &=& \mathcal{G}\left(0.0, 0.1\right).~\label{eq:alpha_d_Planck}
\end{eqnarray} 
In practice, this prior would have to be estimated from simulations or from high-resolution estimations of the foregrounds SEDs. 
The prior, Eq.~\ref{eq:alpha_d_Planck}, is then added to the likelihood defined in Eq.~\ref{eq:spectral_likelihood}. We also update the computation of the analytical error bars on spectral parameters by taking $\boldsymbol{\Sigma}^{-1} \rightarrow \boldsymbol{\Sigma}^{-1} + 1/\sigma^2_\Pi$, where $\sigma_\Pi$ is the width of the Gaussian in Eq.~\ref{eq:alpha_d_Planck}. The resulting foregrounds residuals as well as their modeling  are shown on the right panel of Fig.~\ref{fig:power_spectra_figure_complex_sky}, as the deep orange curves. Despite an increase of their amplitudes, these two residuals now agree quite well. Beyond angular power spectra, this agreement can also be seen in Fig.~\ref{fig:likelihood_complex_sky_sdrun}, where the estimation of $r$ is now essentially unbiased and close to the fiducial case.\\

In \textbf{sim\#2}, the parametric form of the modeled SEDs do not match what is assumed in the input simulations. Among other differences, there is polarized AME in the sky whereas the model does not assume its existence. When estimating the usual set of spectral indices, $\boldsymbol{\beta}\,\in\,\{\beta_d, T_d,\beta_s\}$, the maximum of the spectral likelihood is therefore biased. As described in~\citep{Stompor2016}, this generates systematic residuals, in addition to the statistical ones described in the previous section. The systematic effect is illustrated by the turquoise curve in the right panel of Fig.~\ref{fig:power_spectra_figure_complex_sky}. It shows a significant spurious power at the largest angular scales.

This rise of power at low-$\ell$ cannot be fully mitigated by including $\alpha_d$, as shown as the turquoise likelihood in Fig.~\ref{fig:likelihood_complex_sky_sdrun}: the bias on $r$ turns out to be significantly high, $\sim 0.01$ with an uncertainty an order of magnitude smaller. Considering a higher order moment expansion has been looked at, but requires either a higher sensitivity than what is provided by the assumed instrumental specifications, or the addition of priors on spectral parameters --- and, in particular, on the additional moments. Finding these latter would require numerous foregrounds simulations that has not been explored in this study, but which should lead to similar performance than what is proposed in the next paragraph.\\
\begin{figure*}
	\centering
        \includegraphics[width=0.9\textwidth]{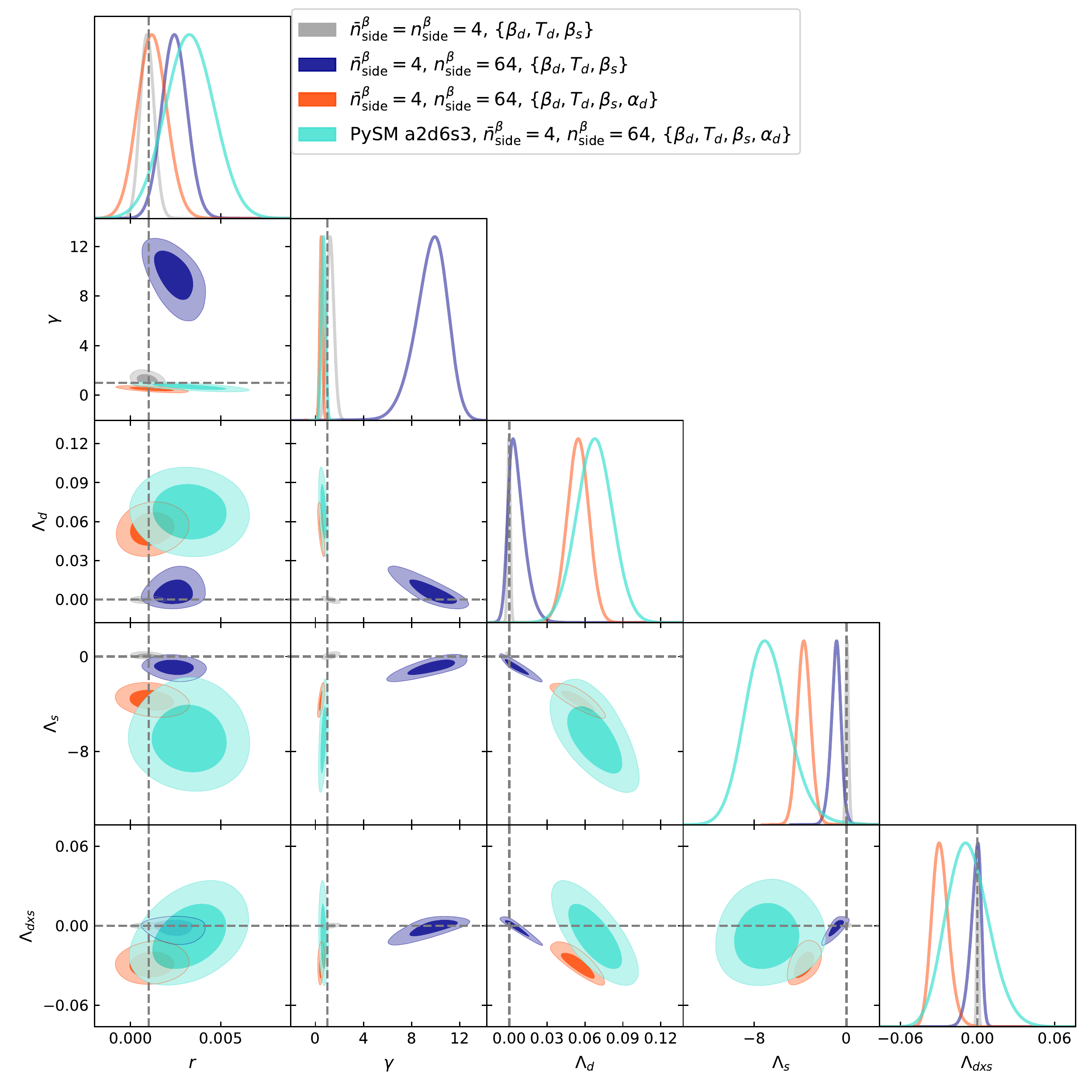}
  \caption{Likelihoods, $\mathcal{L}_{\rm cosmo}(r, \gamma,\boldsymbol{\Lambda}_{i\in\{d,s,d\times s\}})$, and their marginals obtained for the various foregrounds simulations described in Section~\ref{ssec:realistic_case} as recovered with help of a sampling procedure. Gray, dark blue, deep orange and turquoise constraints are respectively for the fiducial case, \textbf{sim\#1} case when fitting for $\{\beta_d,\,\beta_s,\,T_d\}$,  \textbf{sim\#1} case when fitting for $\{\beta_d,\,\beta_s,\,T_d\}$ and a running index in the dust SED, and \textbf{sim\#2} case.}
	\label{fig:likelihood_complex_sky}
\end{figure*}

To reduce this bias on tensor-to-scalar ratio, we explore here the possibility of marginalizing the likelihood on $r$ over templates of foregrounds that are by-products of the proposed component separation algorithm, see Fig.~\ref{fig:explanatory_schematic}.
We propose to update the expression for the modeled covariance of the estimated CMB map, Eq.~\ref{eq:C_ell_def}, with a new term $C_\ell^{\boldsymbol{\Lambda}}$ and a new scalar parameter $\gamma$ such that\footnote{Similarly to Eq.~41 of~\citep{Stompor2016}, we can update the total covariance, Eq.~\ref{eq:C_ell_def}, as:
\begin{eqnarray}
	\centering
    	\mathbf{C}_{jj'} = \mathbf{C}^{\rm CMB+noise}_\ell \delta_{jj'} + \boldsymbol{\Xi}_{jj'} +  \boldsymbol{\Xi}^{\rm sys}_{jj'}
	\label{eq:C_ell_def_sys}
\end{eqnarray}
where $\boldsymbol{\Xi}^{\rm sys}_{jj'}$ is the Fourier representation of $\langle s^T \boldsymbol{\Lambda} s\rangle_{\rm CMB+noise}$, with $\boldsymbol{\Lambda}$ being a free, symmetric, $n_{\rm comp}\,\times\,n_{\rm comp}$ matrix, parametrizing the leakage of foregrounds templates to the final, cleaned CMB map.
},
\begin{eqnarray}
\mathbf{C}_\ell \equiv C_\ell^{\rm prim,\, BB}(r) + C_\ell^{\rm len,\, BB} + N_\ell + \gamma  C_\ell^{\boldsymbol{\Sigma}} +  C_\ell^{\boldsymbol{\Lambda}}
\end{eqnarray}
where,
 \begin{eqnarray}
 	\centering
		\hspace{-0.5cm}C_\ell^{\boldsymbol{\Lambda}} &\equiv& {\sum_{i,j} \boldsymbol{\Lambda}_{ij} \tilde{C}_\ell^{(i\times j)}}\\
		\hspace{-0.5cm}&=& \Lambda_d \tilde{C}_\ell^{\rm (d\times d)} +  \Lambda_s \tilde{C}_\ell^{\rm (s\times s)}  + \Lambda_{d\times s} \tilde{C}_\ell^{\rm (d \times s)}.
	\label{eq:modeling_of_Cl_sys}
\end{eqnarray}
Here, $\tilde{C}_\ell^{(i,j)}$ denotes the  $n_{\rm comp} \times n_{\rm comp}\times \ell_{\rm max}$ angular cross power spectra, estimated from the reconstructed nearly-full sky components $\tilde{s}$, Eq.~\ref{eq:sky_signal_estimate}. The ${\rm d\times d}$, ${\rm s\times s}$, and ${\rm d\times s}$ indices correspond to the dust, synchrotron and dust $\times$ synchrotron auto- and cross-power spectra. 
$\boldsymbol{\Lambda}$ is assumed to be a symmetric $n_{\rm comp} \times n_{\rm comp}$ matrix whose elements are let free. 
We can therefore update the model for $\mathbf{C}_\ell$, Eq.~\ref{eq:C_ell_def}, by adding $\gamma$ and  $\boldsymbol{\Lambda}$ as free parameters, so that the cosmological likelihood on $r$ can be re-expressed as $\mathcal{L}_{\rm cosmo}(r) \rightarrow \mathcal{L}_{\rm cosmo}(r,\gamma, \boldsymbol{\Lambda}_{i\in\{d,s,d\times s\}})$. 
The constraints on the tensor-to-scalar ratio can now be marginalized over the amplitude of boosted statistical foregrounds residuals, through $\gamma$, as well as the modeled systematic foregrounds residuals with $\boldsymbol{\Lambda}$. Fig.~\ref{fig:likelihood_complex_sky} shows the sampling of $\mathcal{L}_{\rm cosmo}(r, \gamma, \boldsymbol{\Lambda}_{i\in\{d,s,d\times s\}})$ using the average of the simulated foregrounds residuals power spectra taken from Fig.~\ref{fig:power_spectra_figure_complex_sky}.

The gray contours correspond to the fiducial case, $\bar{n}^\beta_{\rm side} = n^\beta_{\rm side} = 4$, for which we already showed $\mathcal{L}_{\rm cosmo}(r)$ in Fig.~\ref{fig:likelihoods_ideal_case}. In this case, the fitted $\gamma$ is compatible with $1$ and the elements of $\boldsymbol{\Lambda}$ are all compatible with zero. The recovered $r$ is, as previously found, compatible with the input $r=0.001$. 

The dark blue contours, corresponding to the case  with $\{n_{\rm side}^\beta,\bar{n}_{\rm side}^\beta\} = \{ 64,4\}$, for which we keep on adjusting $\{\beta_d,\,\beta_s,\,T_d\}$ in each sky patch, shows a slightly, although not very significant, bias on $r=0.001$. Yet, through the marginalization of the cosmological likelihood over $\gamma$ and $\boldsymbol{\Lambda}$ parameters, the bias is much smaller than what is observed in Fig.~\ref{fig:likelihood_complex_sky_sdrun}. The marginalization results in a best fit $\gamma > 1$, as it could have been expected from the left panel of Fig.~\ref{fig:power_spectra_figure_complex_sky}. This case shows the consequence of a mismatch between the ``true'' patches and the ones used for the analysis, and how the resulting systematic residuals can be mitigated through the proposed marginalization at the level of angular spectra. 

The deep orange contours, corresponding to the case with $\{n_{\rm side}^\beta,\bar{n}_{\rm side}^\beta\} = \{ 64,4\}$ where we include a running index in the dust SED, $\alpha_d$, is also compatible with the input $r=0.001$, as it was already observed in Fig.~\ref{fig:power_spectra_figure_complex_sky}. Yet, one can see that the best fit shows some nonzero $\boldsymbol{\Lambda}$ elements, which betrays possible degeneracies sourced by the addition of these new free parameters. This case shows that moment expansion is an interesting way to handle the averaging effect, and to keep the modeling of statistical foregrounds residuals valid, even in the case of a mismatch between the ``true'' and the analysis patches.

Finally, the case of complex foregrounds simulations benefits from this marginalization and shows an $r$ which is now compatible with $r=0.001$ (compared to the $\sim$10-$\sigma$ bias in Fig.~\ref{fig:likelihood_complex_sky_sdrun}): the turquoise contours, obtained with the ``a2d6s3'' PySM simulations, show that our model describes relatively well the foregrounds residuals. The best fit shows $\boldsymbol{\Lambda}_{i\in\{d,s,d\times s\}} \neq 0$, $\gamma \neq 1$, and results in a slightly, though statistically not significantly, biased estimate of $r$, along with an expected increased uncertainty compared to the fiducial case.

In practice, for a given instrumental configuration and a given definition of the sky patches, we suggest to proceed in two steps: 1) fit for running indices up to the highest order achievable. This will control the consequences of averaging the non-linear SEDs within the patches. The number of these extra parameters will have to be estimated by looking at e.g. the condition of the $\boldsymbol{\Sigma}$ matrix and the consequences of possible degeneracies with other spectral parameters; 2) marginalize the cosmological likelihood over $\gamma$ and $\boldsymbol{\Lambda}$ parameters, in order to correct for possible systematic leakages of foregrounds into the final cleaned CMB map.

%%%%%%%%%%%%%%%%%%%%%%%%%%%%%%%%%%%%
\section{Conclusions}
\label{sec:conclusions}

In this work, we investigate the performance of the parametric component separation method in the presence of the spatial variability of dust and synchrotron SEDs in the context of future nearly-full sky, satellite-like experimental effort targeting a reliable measurement of the tensor-to-scalar ratio, $r$, as low as $r=0.001$.  We find that using the simplest rendition of the parametric approach, with a single set of spectral parameters used to describe the foregrounds over the entire observed sky, can lead to a spurious detection of value of $r$ as high as $0.01$. Consequently, we study a straightforward extension of the method, which includes independent sets of scaling parameters, $\boldsymbol{\beta}$, assigned to different, disjoint regions of the observed sky.

We show that in the cases when the adapted regions capture well the actual variability of foreground properties then, given the assumed sensitivities, we can robustly recover the values of $r$ as low as the target value of $r=0.001$, but only if the presence of the so-called statistical foreground residuals in the CMB maps and unavoidably arising as a result of the component separation procedure is taken into account. This can be done by introducing an additional term to the covariance matrix of the estimated CMB map.
In contrast, neglecting this term can lead to bias as high as an order of magnitude higher than the true value and the estimated statistical uncertainty.
In a consistent likelihood approach, where both the sky components, $\mathbf{s}$, and the spectral parameters, $\boldsymbol{\beta}$, are considered simultaneously, the extra variance arises naturally as a result of a marginalization over the foreground degrees of freedom, but it has to be introduced manually if the tensor-to-scalar ratio is estimated from the CMB map directly. We propose a generalization of the semianalytic model introduced in~\citep{Stompor2016}, which suitable for the multipatch analysis, and discuss how the correction term can be computed from the available data. 

We also study the angular power spectra of these residuals and elucidate their dependence on the assumptions and the underlying sky model. We show that on angular scales larger that a typical size of the sky patches the statistical residuals are decorrelated and their spectra follow a white noise spectrum. This is due to the fact that the values of $\boldsymbol{\beta}$ estimated for different sky patches are uncorrelated. On smaller angular scales, the spectra have a peak at the scale corresponding to the typical patch size and start decaying at a typical rate of $C_\ell^{res} \propto \ell^{-2.4-2.6}$ on still smaller angular scales. The amplitude of the decaying part is proportional to the input foregrounds spectrum, with the coefficient of proportionality depending on the error bars, $\boldsymbol{\Sigma}$, on the recovered spectral indices, which for a fixed total observed sky area are in turn typically proportional to the number of independent patches, $\boldsymbol{\Sigma} \propto (\bar{n}_{\rm side}^\beta)^2$.

In more complex cases, 
the residuals will also arise due to systematic discrepancies between the assumed and underlying sky models, whenever the component separation model does not have enough flexibility to reflect the reality of foregrounds emissions with sufficient precision. For example, this may be the case when a new, and unaccounted for, sky component is present, such as polarized AME, or the typical scale of the true scaling parameters, $\boldsymbol{\beta}$, is smaller than the considered patch size. 
We illustrate this type of residuals with a new set of foregrounds simulations, assuming that spatial variations of $\boldsymbol{\beta}$ occur on subdegree scales whereas we keep on analyzing SEDs over a $2\leq \bar{n}_{\rm side}^\beta\leq 8$ \textsc{Healpix} grid. In this case, we 
show that adding a free running index to the modeled dust SED
captures the averaging of the multiple gray body emissions over the analysis patch size. This reduces the systematic foregrounds residuals while slightly boosting the statistical ones. We note that the limited signal-to-noise ratio to constrain the dust running index forces us to add a prior on the spectral likelihood. In practice, this latter would be coming from theory and simulations, or from dedicated, sensitive, low- and high-frequency measurements. \\
We finally looked at extended simulations with polarized AME, complex dust model and curvature in the synchrotron emission, while keeping the setup for the component separation unchanged. This case shows significant leakage of foregrounds power to the final CMB map, at all angular scales. To mitigate these systematic residuals, we propose an extra term to the cosmological likelihood which contains 4 new parameters: a multiplicative factor on the statistical foregrounds residuals, and 3 parameters corresponding to the leakage coefficients applied to the recovered full-sky dust and synchrotron auto- and cross-spectra. We conclude based on 10 CMB and noise simulations, that this approach turns out to significantly reduce the bias on the recovered tensor-to-scalar ratio. \\

We emphasize the fact that we have not used any priors on $\boldsymbol{\beta}$ while optimizing the spectral likelihood, Eq.~\ref{eq:estimation_of_beta}, except when including a free running index in the dust SED. Such priors can in principle significantly improve the estimation of $\boldsymbol{\beta}$ and therefore reduce the power of the final foregrounds residuals. This external information would have to be chosen carefully, certainly spatially varying, in order to capture the full complexity of the SEDs across the sky. \\

In practice, several tests could be designed to probe the presence of statistical or systematic contaminants in real datasets:
\begin{itemize}
	\item note that statistical foregrounds residuals, at the power spectrum domain, do not depend on the amplitude of input foregrounds signals~\cite{Errard2011}:
		\begin{eqnarray}
			\centering
				\boldsymbol{\Sigma} &\propto& |s_p|^{-2},\ {\rm cf.\ Eq.~\ref{eqn:secDervSat}} \nonumber \\
				\Rightarrow C_\ell^{\boldsymbol{\Sigma}} &\propto& \boldsymbol{\Sigma}\times |s_p|^2 \propto {\rm constant}
		\end{eqnarray}
		The statistical foregrounds residuals also seem to have a white-noise power spectrum on the largest angular scales, which can \textit{a priori} be disentangled from the primordial $B$-modes theoretical template. This is true as long as the instrumental noise is also decorrelated from one sky patch to another.
	\item on the contrary, systematic foregrounds residuals are leakages from the input foregrounds to the cleaned $B$-modes map i.e.
	\begin{eqnarray}
			\centering
				C_\ell^{\rm sys\, res} &\propto& |s_p|^2
		\end{eqnarray}
	Looking for nonisotropic~\cite{2016JCAP...09..034R} and non-Gaussian signature in the final $B$-modes map, or for any deviation from the black body spectrum are post-processing checks which could be of interest in front of real data sets.
\end{itemize}
The behaviors of these residuals as a function of the sky signals, $s_p$, can in principle be probed by running independently the component separation on different regions of the sky. \\

Finally, this paper describes possible ways to mitigate statistical and systematic foregrounds residuals in the final CMB $B$-modes map, assuming a predefined set of sky patches. 
In reality, properties of foregrounds SEDs will not follow a simple \textsc{Healpix} grid with e.g. $n^\beta_{\rm side}=8$. A way to move forward would be to adaptively characterize and optimize the sizes and shapes of the patches used for the component separation, given the estimated spatial variations of $\boldsymbol{\beta}$ (sourcing the systematic errors and residuals) and given the associated statistical error bars $\boldsymbol{\Sigma}$ (sourcing the statistical errors and residuals). For fixed instrumental specifications, it should be in principle possible to construct patches for each $\boldsymbol{\beta}$ independently, and therefore reduce the statistical foregrounds residuals while keeping the systematic ones under control. This approach will be explored in a later work. \\

\acknowledgements{The authors acknowledge support of the French National Research Agency (Agence National de Recherche) grant, ANR B$\times$B, and thank Thomas Montandon and Baptiste Jost who performed their master internship working on related topics. We also thank Eiichiro Komatsu, Davide Poletti and Carlo Baccigalupi for their useful comments. We acknowledge the use of {\textsc CAMB}\footnote{\url{camb.info}} \cite{camb-1, camb-2}, {\textsc GetDist}\footnote{\url{https://github.com/cmbant/getdist}} and {\textsc Healpix}\footnote{\url{http://healpix.sourceforge.net/}} \cite{healpix} software packages. This work was motivated by discussions within the LiteBIRD\footnote{\url{http://litebird.jp/eng/}} Joint Study Group on foregrounds. Finally, the authors would like to thank the Labex Univearths\footnote{\url{http://www.univearths.fr/fr/le-labex-univearths/}} and CNES\footnote{\url{https://cnes.fr/fr}} for travel support.}

%%%%%%%%%%%%%%%%%%%%%%%%%%%%%%%%%%%%
\bibliographystyle{apsrev}
\bibliography{multipatch}

\begin{thebibliography}{47}
\expandafter\ifx\csname natexlab\endcsname\relax\def\natexlab#1{#1}\fi
\expandafter\ifx\csname bibnamefont\endcsname\relax
  \def\bibnamefont#1{#1}\fi
\expandafter\ifx\csname bibfnamefont\endcsname\relax
  \def\bibfnamefont#1{#1}\fi
\expandafter\ifx\csname citenamefont\endcsname\relax
  \def\citenamefont#1{#1}\fi
\expandafter\ifx\csname url\endcsname\relax
  \def\url#1{\texttt{#1}}\fi
\expandafter\ifx\csname urlprefix\endcsname\relax\def\urlprefix{URL }\fi
\providecommand{\bibinfo}[2]{#2}
\providecommand{\eprint}[2][]{\url{#2}}

\bibitem[{\citenamefont{{Stompor} et~al.}(2016)\citenamefont{{Stompor},
  {Errard}, and {Poletti}}}]{Stompor2016}
\bibinfo{author}{\bibfnamefont{R.}~\bibnamefont{{Stompor}}},
  \bibinfo{author}{\bibfnamefont{J.}~\bibnamefont{{Errard}}}, \bibnamefont{and}
  \bibinfo{author}{\bibfnamefont{D.}~\bibnamefont{{Poletti}}},
  \bibinfo{journal}{\prd} \textbf{\bibinfo{volume}{94}}, \bibinfo{eid}{083526}
  (\bibinfo{year}{2016}), \eprint{1609.03807}.

\bibitem[{\citenamefont{{POLARBEAR Collaboration}
  et~al.}(2017)\citenamefont{{POLARBEAR Collaboration}, {Ade}, {Aguilar},
  {Akiba}, {Arnold}, {Baccigalupi}, {Barron}, {Beck}, {Bianchini}, {Boettger}
  et~al.}}]{2017ApJ...848..121P}
\bibinfo{author}{\bibnamefont{{POLARBEAR Collaboration}}},
  \bibinfo{author}{\bibfnamefont{P.~A.~R.} \bibnamefont{{Ade}}},
  \bibinfo{author}{\bibfnamefont{M.}~\bibnamefont{{Aguilar}}},
  \bibinfo{author}{\bibfnamefont{Y.}~\bibnamefont{{Akiba}}},
  \bibinfo{author}{\bibfnamefont{K.}~\bibnamefont{{Arnold}}},
  \bibinfo{author}{\bibfnamefont{C.}~\bibnamefont{{Baccigalupi}}},
  \bibinfo{author}{\bibfnamefont{D.}~\bibnamefont{{Barron}}},
  \bibinfo{author}{\bibfnamefont{D.}~\bibnamefont{{Beck}}},
  \bibinfo{author}{\bibfnamefont{F.}~\bibnamefont{{Bianchini}}},
  \bibinfo{author}{\bibfnamefont{D.}~\bibnamefont{{Boettger}}},
  \bibnamefont{et~al.}, \bibinfo{journal}{\apj} \textbf{\bibinfo{volume}{848}},
  \bibinfo{eid}{121} (\bibinfo{year}{2017}), \eprint{1705.02907}.

\bibitem[{\citenamefont{{The BICEP/Keck Collaboration}
  et~al.}(2018)\citenamefont{{The BICEP/Keck Collaboration}, {:}, {Ade},
  {Ahmed}, {Aikin}, {Alexander}, {Barkats}, {Benton}, {Bischoff}, {Bock}
  et~al.}}]{2018arXiv180702199T}
\bibinfo{author}{\bibnamefont{{The BICEP/Keck Collaboration}}},
  \bibinfo{author}{\bibnamefont{{:}}}, \bibinfo{author}{\bibfnamefont{P.~A.~R.}
  \bibnamefont{{Ade}}},
  \bibinfo{author}{\bibfnamefont{Z.}~\bibnamefont{{Ahmed}}},
  \bibinfo{author}{\bibfnamefont{R.~W.} \bibnamefont{{Aikin}}},
  \bibinfo{author}{\bibfnamefont{K.~D.} \bibnamefont{{Alexander}}},
  \bibinfo{author}{\bibfnamefont{D.}~\bibnamefont{{Barkats}}},
  \bibinfo{author}{\bibfnamefont{S.~J.} \bibnamefont{{Benton}}},
  \bibinfo{author}{\bibfnamefont{C.~A.} \bibnamefont{{Bischoff}}},
  \bibinfo{author}{\bibfnamefont{J.~J.} \bibnamefont{{Bock}}},
  \bibnamefont{et~al.}, \bibinfo{journal}{ArXiv e-prints}
  (\bibinfo{year}{2018}), \eprint{1807.02199}.

\bibitem[{\citenamefont{{Henning} et~al.}(2018)\citenamefont{{Henning},
  {Sayre}, {Reichardt}, {Ade}, {Anderson}, {Austermann}, {Beall}, {Bender},
  {Benson}, {Bleem} et~al.}}]{2018ApJ...852...97H}
\bibinfo{author}{\bibfnamefont{J.~W.} \bibnamefont{{Henning}}},
  \bibinfo{author}{\bibfnamefont{J.~T.} \bibnamefont{{Sayre}}},
  \bibinfo{author}{\bibfnamefont{C.~L.} \bibnamefont{{Reichardt}}},
  \bibinfo{author}{\bibfnamefont{P.~A.~R.} \bibnamefont{{Ade}}},
  \bibinfo{author}{\bibfnamefont{A.~J.} \bibnamefont{{Anderson}}},
  \bibinfo{author}{\bibfnamefont{J.~E.} \bibnamefont{{Austermann}}},
  \bibinfo{author}{\bibfnamefont{J.~A.} \bibnamefont{{Beall}}},
  \bibinfo{author}{\bibfnamefont{A.~N.} \bibnamefont{{Bender}}},
  \bibinfo{author}{\bibfnamefont{B.~A.} \bibnamefont{{Benson}}},
  \bibinfo{author}{\bibfnamefont{L.~E.} \bibnamefont{{Bleem}}},
  \bibnamefont{et~al.}, \bibinfo{journal}{\apj} \textbf{\bibinfo{volume}{852}},
  \bibinfo{eid}{97} (\bibinfo{year}{2018}), \eprint{1707.09353}.

\bibitem[{\citenamefont{{Louis} et~al.}(2017)\citenamefont{{Louis}, {Grace},
  {Hasselfield}, {Lungu}, {Maurin}, {Addison}, {Ade}, {Aiola}, {Allison},
  {Amiri} et~al.}}]{2017JCAP...06..031L}
\bibinfo{author}{\bibfnamefont{T.}~\bibnamefont{{Louis}}},
  \bibinfo{author}{\bibfnamefont{E.}~\bibnamefont{{Grace}}},
  \bibinfo{author}{\bibfnamefont{M.}~\bibnamefont{{Hasselfield}}},
  \bibinfo{author}{\bibfnamefont{M.}~\bibnamefont{{Lungu}}},
  \bibinfo{author}{\bibfnamefont{L.}~\bibnamefont{{Maurin}}},
  \bibinfo{author}{\bibfnamefont{G.~E.} \bibnamefont{{Addison}}},
  \bibinfo{author}{\bibfnamefont{P.~A.~R.} \bibnamefont{{Ade}}},
  \bibinfo{author}{\bibfnamefont{S.}~\bibnamefont{{Aiola}}},
  \bibinfo{author}{\bibfnamefont{R.}~\bibnamefont{{Allison}}},
  \bibinfo{author}{\bibfnamefont{M.}~\bibnamefont{{Amiri}}},
  \bibnamefont{et~al.}, \bibinfo{journal}{\jcap} \textbf{\bibinfo{volume}{6}},
  \bibinfo{eid}{031} (\bibinfo{year}{2017}), \eprint{1610.02360}.

\bibitem[{\citenamefont{{Grayson} et~al.}(2016)\citenamefont{{Grayson}, {Ade},
  {Ahmed}, {Alexander}, {Amiri}, {Barkats}, {Benton}, {Bischoff}, {Bock},
  {Boenish} et~al.}}]{2016SPIE.9914E..0SG}
\bibinfo{author}{\bibfnamefont{J.~A.} \bibnamefont{{Grayson}}},
  \bibinfo{author}{\bibfnamefont{P.~A.~R.} \bibnamefont{{Ade}}},
  \bibinfo{author}{\bibfnamefont{Z.}~\bibnamefont{{Ahmed}}},
  \bibinfo{author}{\bibfnamefont{K.~D.} \bibnamefont{{Alexander}}},
  \bibinfo{author}{\bibfnamefont{M.}~\bibnamefont{{Amiri}}},
  \bibinfo{author}{\bibfnamefont{D.}~\bibnamefont{{Barkats}}},
  \bibinfo{author}{\bibfnamefont{S.~J.} \bibnamefont{{Benton}}},
  \bibinfo{author}{\bibfnamefont{C.~A.} \bibnamefont{{Bischoff}}},
  \bibinfo{author}{\bibfnamefont{J.~J.} \bibnamefont{{Bock}}},
  \bibinfo{author}{\bibfnamefont{H.}~\bibnamefont{{Boenish}}},
  \bibnamefont{et~al.} (\bibinfo{year}{2016}), vol. \bibinfo{volume}{9914} of
  \emph{\bibinfo{series}{\procspie}}, p. \bibinfo{pages}{99140S},
  \eprint{1607.04668}.

\bibitem[{\citenamefont{{Barron} and {POLARBEAR
  Collaboration}}(2018)}]{2018AAS...23135602B}
\bibinfo{author}{\bibfnamefont{D.}~\bibnamefont{{Barron}}} \bibnamefont{and}
  \bibinfo{author}{\bibnamefont{{POLARBEAR Collaboration}}}
  (\bibinfo{year}{2018}), vol. \bibinfo{volume}{231} of
  \emph{\bibinfo{series}{American Astronomical Society Meeting Abstracts}}, p.
  \bibinfo{pages}{356.02}.

\bibitem[{\citenamefont{{Choi} et~al.}(2018)\citenamefont{{Choi}, {Austermann},
  {Beall}, {Crowley}, {Datta}, {Duff}, {Gallardo}, {Ho}, {Hubmayr}, {Koopman}
  et~al.}}]{2018JLTP..tmp..159C}
\bibinfo{author}{\bibfnamefont{S.~K.} \bibnamefont{{Choi}}},
  \bibinfo{author}{\bibfnamefont{J.}~\bibnamefont{{Austermann}}},
  \bibinfo{author}{\bibfnamefont{J.~A.} \bibnamefont{{Beall}}},
  \bibinfo{author}{\bibfnamefont{K.~T.} \bibnamefont{{Crowley}}},
  \bibinfo{author}{\bibfnamefont{R.}~\bibnamefont{{Datta}}},
  \bibinfo{author}{\bibfnamefont{S.~M.} \bibnamefont{{Duff}}},
  \bibinfo{author}{\bibfnamefont{P.~A.} \bibnamefont{{Gallardo}}},
  \bibinfo{author}{\bibfnamefont{S.~P.} \bibnamefont{{Ho}}},
  \bibinfo{author}{\bibfnamefont{J.}~\bibnamefont{{Hubmayr}}},
  \bibinfo{author}{\bibfnamefont{B.~J.} \bibnamefont{{Koopman}}},
  \bibnamefont{et~al.}, \bibinfo{journal}{Journal of Low Temperature Physics}
  (\bibinfo{year}{2018}), \eprint{1711.04841}.

\bibitem[{\citenamefont{{Sobrin} et~al.}(2018)\citenamefont{{Sobrin}, {Ade},
  {Ahmed}, {Anderson}, {Avva}, {Basu Thakur}, {Bender}, {Benson}, {Carlstrom},
  {Carter} et~al.}}]{2018arXiv180900032S}
\bibinfo{author}{\bibfnamefont{J.~A.} \bibnamefont{{Sobrin}}},
  \bibinfo{author}{\bibfnamefont{P.~A.~R.} \bibnamefont{{Ade}}},
  \bibinfo{author}{\bibfnamefont{Z.}~\bibnamefont{{Ahmed}}},
  \bibinfo{author}{\bibfnamefont{A.~J.} \bibnamefont{{Anderson}}},
  \bibinfo{author}{\bibfnamefont{J.~S.} \bibnamefont{{Avva}}},
  \bibinfo{author}{\bibfnamefont{R.}~\bibnamefont{{Basu Thakur}}},
  \bibinfo{author}{\bibfnamefont{A.~N.} \bibnamefont{{Bender}}},
  \bibinfo{author}{\bibfnamefont{B.~A.} \bibnamefont{{Benson}}},
  \bibinfo{author}{\bibfnamefont{J.~E.} \bibnamefont{{Carlstrom}}},
  \bibinfo{author}{\bibfnamefont{F.~W.} \bibnamefont{{Carter}}},
  \bibnamefont{et~al.}, \bibinfo{journal}{ArXiv e-prints}
  (\bibinfo{year}{2018}), \eprint{1809.00032}.

\bibitem[{\citenamefont{{The Simons Observatory Collaboration}
  et~al.}(2018)\citenamefont{{The Simons Observatory Collaboration}, {Ade},
  {Aguirre}, {Ahmed}, {Aiola}, {Ali}, {Alonso}, {Alvarez}, {Arnold}, {Ashton}
  et~al.}}]{Simons_Observatory}
\bibinfo{author}{\bibnamefont{{The Simons Observatory Collaboration}}},
  \bibinfo{author}{\bibfnamefont{P.}~\bibnamefont{{Ade}}},
  \bibinfo{author}{\bibfnamefont{J.}~\bibnamefont{{Aguirre}}},
  \bibinfo{author}{\bibfnamefont{Z.}~\bibnamefont{{Ahmed}}},
  \bibinfo{author}{\bibfnamefont{S.}~\bibnamefont{{Aiola}}},
  \bibinfo{author}{\bibfnamefont{A.}~\bibnamefont{{Ali}}},
  \bibinfo{author}{\bibfnamefont{D.}~\bibnamefont{{Alonso}}},
  \bibinfo{author}{\bibfnamefont{M.~A.} \bibnamefont{{Alvarez}}},
  \bibinfo{author}{\bibfnamefont{K.}~\bibnamefont{{Arnold}}},
  \bibinfo{author}{\bibfnamefont{P.}~\bibnamefont{{Ashton}}},
  \bibnamefont{et~al.}, \bibinfo{journal}{ArXiv e-prints}
  (\bibinfo{year}{2018}), \eprint{1808.07445}.

\bibitem[{\citenamefont{{Hui} et~al.}(2018)\citenamefont{{Hui}, {Ade}, {Ahmed},
  {Aikin}, {Alexander}, {Barkats}, {Benton}, {Bischoff}, {Bock}, {Bowens-Rubin}
  et~al.}}]{2018arXiv180800568H}
\bibinfo{author}{\bibfnamefont{H.}~\bibnamefont{{Hui}}},
  \bibinfo{author}{\bibfnamefont{P.~A.~R.} \bibnamefont{{Ade}}},
  \bibinfo{author}{\bibfnamefont{Z.}~\bibnamefont{{Ahmed}}},
  \bibinfo{author}{\bibfnamefont{R.~W.} \bibnamefont{{Aikin}}},
  \bibinfo{author}{\bibfnamefont{K.~D.} \bibnamefont{{Alexander}}},
  \bibinfo{author}{\bibfnamefont{D.}~\bibnamefont{{Barkats}}},
  \bibinfo{author}{\bibfnamefont{S.~J.} \bibnamefont{{Benton}}},
  \bibinfo{author}{\bibfnamefont{C.~A.} \bibnamefont{{Bischoff}}},
  \bibinfo{author}{\bibfnamefont{J.~J.} \bibnamefont{{Bock}}},
  \bibinfo{author}{\bibfnamefont{R.}~\bibnamefont{{Bowens-Rubin}}},
  \bibnamefont{et~al.}, \bibinfo{journal}{ArXiv e-prints}
  (\bibinfo{year}{2018}), \eprint{1808.00568}.

\bibitem[{\citenamefont{{Abazajian} et~al.}(2016)\citenamefont{{Abazajian},
  {Adshead}, {Ahmed}, {Allen}, {Alonso}, {Arnold}, {Baccigalupi}, {Bartlett},
  {Battaglia}, {Benson} et~al.}}]{2016arXiv161002743A}
\bibinfo{author}{\bibfnamefont{K.~N.} \bibnamefont{{Abazajian}}},
  \bibinfo{author}{\bibfnamefont{P.}~\bibnamefont{{Adshead}}},
  \bibinfo{author}{\bibfnamefont{Z.}~\bibnamefont{{Ahmed}}},
  \bibinfo{author}{\bibfnamefont{S.~W.} \bibnamefont{{Allen}}},
  \bibinfo{author}{\bibfnamefont{D.}~\bibnamefont{{Alonso}}},
  \bibinfo{author}{\bibfnamefont{K.~S.} \bibnamefont{{Arnold}}},
  \bibinfo{author}{\bibfnamefont{C.}~\bibnamefont{{Baccigalupi}}},
  \bibinfo{author}{\bibfnamefont{J.~G.} \bibnamefont{{Bartlett}}},
  \bibinfo{author}{\bibfnamefont{N.}~\bibnamefont{{Battaglia}}},
  \bibinfo{author}{\bibfnamefont{B.~A.} \bibnamefont{{Benson}}},
  \bibnamefont{et~al.}, \bibinfo{journal}{ArXiv e-prints}
  (\bibinfo{year}{2016}), \eprint{1610.02743}.

\bibitem[{\citenamefont{{Matsumura} et~al.}(2016)\citenamefont{{Matsumura},
  {Akiba}, {Arnold}, {Borrill}, {Chendra}, {Chinone}, {Cukierman}, {de Haan},
  {Dobbs}, {Dominjon} et~al.}}]{2016JLTP..184..824M}
\bibinfo{author}{\bibfnamefont{T.}~\bibnamefont{{Matsumura}}},
  \bibinfo{author}{\bibfnamefont{Y.}~\bibnamefont{{Akiba}}},
  \bibinfo{author}{\bibfnamefont{K.}~\bibnamefont{{Arnold}}},
  \bibinfo{author}{\bibfnamefont{J.}~\bibnamefont{{Borrill}}},
  \bibinfo{author}{\bibfnamefont{R.}~\bibnamefont{{Chendra}}},
  \bibinfo{author}{\bibfnamefont{Y.}~\bibnamefont{{Chinone}}},
  \bibinfo{author}{\bibfnamefont{A.}~\bibnamefont{{Cukierman}}},
  \bibinfo{author}{\bibfnamefont{T.}~\bibnamefont{{de Haan}}},
  \bibinfo{author}{\bibfnamefont{M.}~\bibnamefont{{Dobbs}}},
  \bibinfo{author}{\bibfnamefont{A.}~\bibnamefont{{Dominjon}}},
  \bibnamefont{et~al.}, \bibinfo{journal}{Journal of Low Temperature Physics}
  \textbf{\bibinfo{volume}{184}}, \bibinfo{pages}{824} (\bibinfo{year}{2016}).

\bibitem[{\citenamefont{{Kogut} et~al.}(2016)\citenamefont{{Kogut}, {Chluba},
  {Fixsen}, {Meyer}, and {Spergel}}}]{2016SPIE.9904E..0WK}
\bibinfo{author}{\bibfnamefont{A.}~\bibnamefont{{Kogut}}},
  \bibinfo{author}{\bibfnamefont{J.}~\bibnamefont{{Chluba}}},
  \bibinfo{author}{\bibfnamefont{D.~J.} \bibnamefont{{Fixsen}}},
  \bibinfo{author}{\bibfnamefont{S.}~\bibnamefont{{Meyer}}}, \bibnamefont{and}
  \bibinfo{author}{\bibfnamefont{D.}~\bibnamefont{{Spergel}}}
  (\bibinfo{year}{2016}), vol. \bibinfo{volume}{9904} of
  \emph{\bibinfo{series}{\procspie}}, p. \bibinfo{pages}{99040W}.

\bibitem[{\citenamefont{{The PICO Collaboration}}(2018)}]{PICO}
\bibinfo{author}{\bibnamefont{{The PICO Collaboration}}},
  \emph{\bibinfo{title}{{CMB Probe Mission Study Wiki}}},
  \bibinfo{howpublished}{\url{https://zzz.physics.umn.edu/ipsig/start}}
  (\bibinfo{year}{2018}).

\bibitem[{\citenamefont{{Lyth}}(1998)}]{1998cosm.work..355L}
\bibinfo{author}{\bibfnamefont{D.~H.} \bibnamefont{{Lyth}}}
  (\bibinfo{year}{1998}), pp. \bibinfo{pages}{355--359}.

\bibitem[{\citenamefont{{Martin}
  et~al.}(2014{\natexlab{a}})\citenamefont{{Martin}, {Ringeval}, and
  {Vennin}}}]{Martin_2014a}
\bibinfo{author}{\bibfnamefont{J.}~\bibnamefont{{Martin}}},
  \bibinfo{author}{\bibfnamefont{C.}~\bibnamefont{{Ringeval}}},
  \bibnamefont{and} \bibinfo{author}{\bibfnamefont{V.}~\bibnamefont{{Vennin}}},
  \bibinfo{journal}{Physics of the Dark Universe} \textbf{\bibinfo{volume}{5}},
  \bibinfo{pages}{75} (\bibinfo{year}{2014}{\natexlab{a}}), \eprint{1303.3787}.

\bibitem[{\citenamefont{{Martin}
  et~al.}(2014{\natexlab{b}})\citenamefont{{Martin}, {Ringeval}, {Trotta}, and
  {Vennin}}}]{Martin_2014b}
\bibinfo{author}{\bibfnamefont{J.}~\bibnamefont{{Martin}}},
  \bibinfo{author}{\bibfnamefont{C.}~\bibnamefont{{Ringeval}}},
  \bibinfo{author}{\bibfnamefont{R.}~\bibnamefont{{Trotta}}}, \bibnamefont{and}
  \bibinfo{author}{\bibfnamefont{V.}~\bibnamefont{{Vennin}}},
  \bibinfo{journal}{\jcap} \textbf{\bibinfo{volume}{3}}, \bibinfo{eid}{039}
  (\bibinfo{year}{2014}{\natexlab{b}}), \eprint{1312.3529}.

\bibitem[{\citenamefont{{Krachmalnicoff}
  et~al.}(2016)\citenamefont{{Krachmalnicoff}, {Baccigalupi}, {Aumont},
  {Bersanelli}, and {Mennella}}}]{Krach2016}
\bibinfo{author}{\bibfnamefont{N.}~\bibnamefont{{Krachmalnicoff}}},
  \bibinfo{author}{\bibfnamefont{C.}~\bibnamefont{{Baccigalupi}}},
  \bibinfo{author}{\bibfnamefont{J.}~\bibnamefont{{Aumont}}},
  \bibinfo{author}{\bibfnamefont{M.}~\bibnamefont{{Bersanelli}}},
  \bibnamefont{and}
  \bibinfo{author}{\bibfnamefont{A.}~\bibnamefont{{Mennella}}},
  \bibinfo{journal}{\aap} \textbf{\bibinfo{volume}{588}}, \bibinfo{eid}{A65}
  (\bibinfo{year}{2016}).

\bibitem[{\citenamefont{{Planck Collaboration}
  et~al.}(2018{\natexlab{a}})\citenamefont{{Planck Collaboration}, {Akrami},
  {Ashdown}, {Aumont}, {Baccigalupi}, {Ballardini}, {Banday}, {Barreiro},
  {Bartolo}, {Basak} et~al.}}]{Planck_LIV_2018}
\bibinfo{author}{\bibnamefont{{Planck Collaboration}}},
  \bibinfo{author}{\bibfnamefont{Y.}~\bibnamefont{{Akrami}}},
  \bibinfo{author}{\bibfnamefont{M.}~\bibnamefont{{Ashdown}}},
  \bibinfo{author}{\bibfnamefont{J.}~\bibnamefont{{Aumont}}},
  \bibinfo{author}{\bibfnamefont{C.}~\bibnamefont{{Baccigalupi}}},
  \bibinfo{author}{\bibfnamefont{M.}~\bibnamefont{{Ballardini}}},
  \bibinfo{author}{\bibfnamefont{A.~J.} \bibnamefont{{Banday}}},
  \bibinfo{author}{\bibfnamefont{R.~B.} \bibnamefont{{Barreiro}}},
  \bibinfo{author}{\bibfnamefont{N.}~\bibnamefont{{Bartolo}}},
  \bibinfo{author}{\bibfnamefont{S.}~\bibnamefont{{Basak}}},
  \bibnamefont{et~al.}, \bibinfo{journal}{ArXiv e-prints}
  (\bibinfo{year}{2018}{\natexlab{a}}), \eprint{1801.04945}.

\bibitem[{\citenamefont{{Planck Collaboration}
  et~al.}(2018{\natexlab{b}})\citenamefont{{Planck Collaboration}, {Aghanim},
  {Akrami}, {Alves}, {Ashdown}, {Aumont}, {Baccigalupi}, {Ballardini},
  {Banday}, {Barreiro} et~al.}}]{2018arXiv180706212P}
\bibinfo{author}{\bibnamefont{{Planck Collaboration}}},
  \bibinfo{author}{\bibfnamefont{N.}~\bibnamefont{{Aghanim}}},
  \bibinfo{author}{\bibfnamefont{Y.}~\bibnamefont{{Akrami}}},
  \bibinfo{author}{\bibfnamefont{M.~I.~R.} \bibnamefont{{Alves}}},
  \bibinfo{author}{\bibfnamefont{M.}~\bibnamefont{{Ashdown}}},
  \bibinfo{author}{\bibfnamefont{J.}~\bibnamefont{{Aumont}}},
  \bibinfo{author}{\bibfnamefont{C.}~\bibnamefont{{Baccigalupi}}},
  \bibinfo{author}{\bibfnamefont{M.}~\bibnamefont{{Ballardini}}},
  \bibinfo{author}{\bibfnamefont{A.~J.} \bibnamefont{{Banday}}},
  \bibinfo{author}{\bibfnamefont{R.~B.} \bibnamefont{{Barreiro}}},
  \bibnamefont{et~al.}, \bibinfo{journal}{ArXiv e-prints}
  (\bibinfo{year}{2018}{\natexlab{b}}), \eprint{1807.06212}.

\bibitem[{\citenamefont{{Planck Collaboration}
  et~al.}(2018{\natexlab{c}})\citenamefont{{Planck Collaboration}, {Akrami},
  {Ashdown}, {Aumont}, {Baccigalupi}, {Ballardini}, {Banday}, {Barreiro},
  {Bartolo}, {Basak} et~al.}}]{2018arXiv180104945P}
\bibinfo{author}{\bibnamefont{{Planck Collaboration}}},
  \bibinfo{author}{\bibfnamefont{Y.}~\bibnamefont{{Akrami}}},
  \bibinfo{author}{\bibfnamefont{M.}~\bibnamefont{{Ashdown}}},
  \bibinfo{author}{\bibfnamefont{J.}~\bibnamefont{{Aumont}}},
  \bibinfo{author}{\bibfnamefont{C.}~\bibnamefont{{Baccigalupi}}},
  \bibinfo{author}{\bibfnamefont{M.}~\bibnamefont{{Ballardini}}},
  \bibinfo{author}{\bibfnamefont{A.~J.} \bibnamefont{{Banday}}},
  \bibinfo{author}{\bibfnamefont{R.~B.} \bibnamefont{{Barreiro}}},
  \bibinfo{author}{\bibfnamefont{N.}~\bibnamefont{{Bartolo}}},
  \bibinfo{author}{\bibfnamefont{S.}~\bibnamefont{{Basak}}},
  \bibnamefont{et~al.}, \bibinfo{journal}{ArXiv e-prints}
  (\bibinfo{year}{2018}{\natexlab{c}}), \eprint{1801.04945}.

\bibitem[{\citenamefont{{Vansyngel} et~al.}(2017)\citenamefont{{Vansyngel},
  {Boulanger}, {Ghosh}, {Wandelt}, {Aumont}, {Bracco}, {Levrier}, {Martin}, and
  {Montier}}}]{Vansyngel2017}
\bibinfo{author}{\bibfnamefont{F.}~\bibnamefont{{Vansyngel}}},
  \bibinfo{author}{\bibfnamefont{F.}~\bibnamefont{{Boulanger}}},
  \bibinfo{author}{\bibfnamefont{T.}~\bibnamefont{{Ghosh}}},
  \bibinfo{author}{\bibfnamefont{B.}~\bibnamefont{{Wandelt}}},
  \bibinfo{author}{\bibfnamefont{J.}~\bibnamefont{{Aumont}}},
  \bibinfo{author}{\bibfnamefont{A.}~\bibnamefont{{Bracco}}},
  \bibinfo{author}{\bibfnamefont{F.}~\bibnamefont{{Levrier}}},
  \bibinfo{author}{\bibfnamefont{P.~G.} \bibnamefont{{Martin}}},
  \bibnamefont{and}
  \bibinfo{author}{\bibfnamefont{L.}~\bibnamefont{{Montier}}},
  \bibinfo{journal}{\aap} \textbf{\bibinfo{volume}{603}}, \bibinfo{eid}{A62}
  (\bibinfo{year}{2017}), \eprint{1611.02577}.

\bibitem[{\citenamefont{{Levrier} et~al.}(2018)\citenamefont{{Levrier},
  {Neveu}, {Falgarone}, {Boulanger}, {Bracco}, {Ghosh}, and
  {Vansyngel}}}]{Levrier_2018}
\bibinfo{author}{\bibfnamefont{F.}~\bibnamefont{{Levrier}}},
  \bibinfo{author}{\bibfnamefont{J.}~\bibnamefont{{Neveu}}},
  \bibinfo{author}{\bibfnamefont{E.}~\bibnamefont{{Falgarone}}},
  \bibinfo{author}{\bibfnamefont{F.}~\bibnamefont{{Boulanger}}},
  \bibinfo{author}{\bibfnamefont{A.}~\bibnamefont{{Bracco}}},
  \bibinfo{author}{\bibfnamefont{T.}~\bibnamefont{{Ghosh}}}, \bibnamefont{and}
  \bibinfo{author}{\bibfnamefont{F.}~\bibnamefont{{Vansyngel}}},
  \bibinfo{journal}{\aap} \textbf{\bibinfo{volume}{614}}, \bibinfo{eid}{A124}
  (\bibinfo{year}{2018}), \eprint{1802.08725}.

\bibitem[{\citenamefont{{Krachmalnicoff}
  et~al.}(2018)\citenamefont{{Krachmalnicoff}, {Carretti}, {Baccigalupi},
  {Bernardi}, {Brown}, {Gaensler}, {Haverkorn}, {Kesteven}, {Perrotta}, {Poppi}
  et~al.}}]{2018arXiv180201145K}
\bibinfo{author}{\bibfnamefont{N.}~\bibnamefont{{Krachmalnicoff}}},
  \bibinfo{author}{\bibfnamefont{E.}~\bibnamefont{{Carretti}}},
  \bibinfo{author}{\bibfnamefont{C.}~\bibnamefont{{Baccigalupi}}},
  \bibinfo{author}{\bibfnamefont{G.}~\bibnamefont{{Bernardi}}},
  \bibinfo{author}{\bibfnamefont{S.}~\bibnamefont{{Brown}}},
  \bibinfo{author}{\bibfnamefont{B.~M.} \bibnamefont{{Gaensler}}},
  \bibinfo{author}{\bibfnamefont{M.}~\bibnamefont{{Haverkorn}}},
  \bibinfo{author}{\bibfnamefont{M.}~\bibnamefont{{Kesteven}}},
  \bibinfo{author}{\bibfnamefont{F.}~\bibnamefont{{Perrotta}}},
  \bibinfo{author}{\bibfnamefont{S.}~\bibnamefont{{Poppi}}},
  \bibnamefont{et~al.}, \bibinfo{journal}{ArXiv e-prints}
  (\bibinfo{year}{2018}), \eprint{1802.01145}.

\bibitem[{\citenamefont{{Planck Collaboration}
  et~al.}(2017)\citenamefont{{Planck Collaboration}, {Aghanim}, {Ashdown},
  {Aumont}, {Baccigalupi}, {Ballardini}, {Banday}, {Barreiro}, {Bartolo},
  {Basak} et~al.}}]{Planck_L_2017}
\bibinfo{author}{\bibnamefont{{Planck Collaboration}}},
  \bibinfo{author}{\bibfnamefont{N.}~\bibnamefont{{Aghanim}}},
  \bibinfo{author}{\bibfnamefont{M.}~\bibnamefont{{Ashdown}}},
  \bibinfo{author}{\bibfnamefont{J.}~\bibnamefont{{Aumont}}},
  \bibinfo{author}{\bibfnamefont{C.}~\bibnamefont{{Baccigalupi}}},
  \bibinfo{author}{\bibfnamefont{M.}~\bibnamefont{{Ballardini}}},
  \bibinfo{author}{\bibfnamefont{A.~J.} \bibnamefont{{Banday}}},
  \bibinfo{author}{\bibfnamefont{R.~B.} \bibnamefont{{Barreiro}}},
  \bibinfo{author}{\bibfnamefont{N.}~\bibnamefont{{Bartolo}}},
  \bibinfo{author}{\bibfnamefont{S.}~\bibnamefont{{Basak}}},
  \bibnamefont{et~al.}, \bibinfo{journal}{\aap} \textbf{\bibinfo{volume}{599}},
  \bibinfo{eid}{A51} (\bibinfo{year}{2017}), \eprint{1606.07335}.

\bibitem[{\citenamefont{{Planck Collaboration}
  et~al.}(2016{\natexlab{a}})\citenamefont{{Planck Collaboration}, {Ade},
  {Aghanim}, {Alves}, {Arnaud}, {Ashdown}, {Aumont}, {Baccigalupi}, {Banday},
  {Barreiro} et~al.}}]{2016AA...594A..25P}
\bibinfo{author}{\bibnamefont{{Planck Collaboration}}},
  \bibinfo{author}{\bibfnamefont{P.~A.~R.} \bibnamefont{{Ade}}},
  \bibinfo{author}{\bibfnamefont{N.}~\bibnamefont{{Aghanim}}},
  \bibinfo{author}{\bibfnamefont{M.~I.~R.} \bibnamefont{{Alves}}},
  \bibinfo{author}{\bibfnamefont{M.}~\bibnamefont{{Arnaud}}},
  \bibinfo{author}{\bibfnamefont{M.}~\bibnamefont{{Ashdown}}},
  \bibinfo{author}{\bibfnamefont{J.}~\bibnamefont{{Aumont}}},
  \bibinfo{author}{\bibfnamefont{C.}~\bibnamefont{{Baccigalupi}}},
  \bibinfo{author}{\bibfnamefont{A.~J.} \bibnamefont{{Banday}}},
  \bibinfo{author}{\bibfnamefont{R.~B.} \bibnamefont{{Barreiro}}},
  \bibnamefont{et~al.}, \bibinfo{journal}{\aap} \textbf{\bibinfo{volume}{594}},
  \bibinfo{eid}{A25} (\bibinfo{year}{2016}{\natexlab{a}}), \eprint{1506.06660}.

\bibitem[{\citenamefont{{Stivoli} et~al.}(2010)\citenamefont{{Stivoli},
  {Grain}, {Leach}, {Tristram}, {Baccigalupi}, and {Stompor}}}]{Stivoli2010}
\bibinfo{author}{\bibfnamefont{F.}~\bibnamefont{{Stivoli}}},
  \bibinfo{author}{\bibfnamefont{J.}~\bibnamefont{{Grain}}},
  \bibinfo{author}{\bibfnamefont{S.~M.} \bibnamefont{{Leach}}},
  \bibinfo{author}{\bibfnamefont{M.}~\bibnamefont{{Tristram}}},
  \bibinfo{author}{\bibfnamefont{C.}~\bibnamefont{{Baccigalupi}}},
  \bibnamefont{and}
  \bibinfo{author}{\bibfnamefont{R.}~\bibnamefont{{Stompor}}},
  \bibinfo{journal}{\mnras} \textbf{\bibinfo{volume}{408}},
  \bibinfo{pages}{2319} (\bibinfo{year}{2010}), \eprint{1004.4756}.

\bibitem[{\citenamefont{{Brandt} et~al.}(1994)\citenamefont{{Brandt},
  {Lawrence}, {Readhead}, {Pakianathan}, and {Fiola}}}]{brandt1994}
\bibinfo{author}{\bibfnamefont{W.~N.} \bibnamefont{{Brandt}}},
  \bibinfo{author}{\bibfnamefont{C.~R.} \bibnamefont{{Lawrence}}},
  \bibinfo{author}{\bibfnamefont{A.~C.~S.} \bibnamefont{{Readhead}}},
  \bibinfo{author}{\bibfnamefont{J.~N.} \bibnamefont{{Pakianathan}}},
  \bibnamefont{and} \bibinfo{author}{\bibfnamefont{T.~M.}
  \bibnamefont{{Fiola}}}, \bibinfo{journal}{\apj}
  \textbf{\bibinfo{volume}{424}}, \bibinfo{pages}{1} (\bibinfo{year}{1994}).

\bibitem[{\citenamefont{{Eriksen} et~al.}(2006)\citenamefont{{Eriksen},
  {Dickinson}, {Lawrence}, {Baccigalupi}, {Banday}, {G{\'o}rski}, {Hansen},
  {Lilje}, {Pierpaoli}, {Seiffert} et~al.}}]{2006ApJ...641..665E}
\bibinfo{author}{\bibfnamefont{H.~K.} \bibnamefont{{Eriksen}}},
  \bibinfo{author}{\bibfnamefont{C.}~\bibnamefont{{Dickinson}}},
  \bibinfo{author}{\bibfnamefont{C.~R.} \bibnamefont{{Lawrence}}},
  \bibinfo{author}{\bibfnamefont{C.}~\bibnamefont{{Baccigalupi}}},
  \bibinfo{author}{\bibfnamefont{A.~J.} \bibnamefont{{Banday}}},
  \bibinfo{author}{\bibfnamefont{K.~M.} \bibnamefont{{G{\'o}rski}}},
  \bibinfo{author}{\bibfnamefont{F.~K.} \bibnamefont{{Hansen}}},
  \bibinfo{author}{\bibfnamefont{P.~B.} \bibnamefont{{Lilje}}},
  \bibinfo{author}{\bibfnamefont{E.}~\bibnamefont{{Pierpaoli}}},
  \bibinfo{author}{\bibfnamefont{M.~D.} \bibnamefont{{Seiffert}}},
  \bibnamefont{et~al.}, \bibinfo{journal}{\apj} \textbf{\bibinfo{volume}{641}},
  \bibinfo{pages}{665} (\bibinfo{year}{2006}), \eprint{arXiv:astro-ph/0508268}.

\bibitem[{\citenamefont{{Planck Collaboration}
  et~al.}(2014)\citenamefont{{Planck Collaboration}, {Abergel}, {Ade},
  {Aghanim}, {Alves}, {Aniano}, {Armitage-Caplan}, {Arnaud}, {Ashdown},
  {Atrio-Barandela} et~al.}}]{Planck_XI_2014}
\bibinfo{author}{\bibnamefont{{Planck Collaboration}}},
  \bibinfo{author}{\bibfnamefont{A.}~\bibnamefont{{Abergel}}},
  \bibinfo{author}{\bibfnamefont{P.~A.~R.} \bibnamefont{{Ade}}},
  \bibinfo{author}{\bibfnamefont{N.}~\bibnamefont{{Aghanim}}},
  \bibinfo{author}{\bibfnamefont{M.~I.~R.} \bibnamefont{{Alves}}},
  \bibinfo{author}{\bibfnamefont{G.}~\bibnamefont{{Aniano}}},
  \bibinfo{author}{\bibfnamefont{C.}~\bibnamefont{{Armitage-Caplan}}},
  \bibinfo{author}{\bibfnamefont{M.}~\bibnamefont{{Arnaud}}},
  \bibinfo{author}{\bibfnamefont{M.}~\bibnamefont{{Ashdown}}},
  \bibinfo{author}{\bibfnamefont{F.}~\bibnamefont{{Atrio-Barandela}}},
  \bibnamefont{et~al.}, \bibinfo{journal}{\aap} \textbf{\bibinfo{volume}{571}},
  \bibinfo{eid}{A11} (\bibinfo{year}{2014}), \eprint{1312.1300}.

\bibitem[{\citenamefont{{Chluba} et~al.}(2017)\citenamefont{{Chluba}, {Hill},
  and {Abitbol}}}]{2017MNRAS.472.1195C}
\bibinfo{author}{\bibfnamefont{J.}~\bibnamefont{{Chluba}}},
  \bibinfo{author}{\bibfnamefont{J.~C.} \bibnamefont{{Hill}}},
  \bibnamefont{and} \bibinfo{author}{\bibfnamefont{M.~H.}
  \bibnamefont{{Abitbol}}}, \bibinfo{journal}{\mnras}
  \textbf{\bibinfo{volume}{472}}, \bibinfo{pages}{1195} (\bibinfo{year}{2017}),
  \eprint{1701.00274}.

\bibitem[{\citenamefont{{G{\'o}rski} et~al.}(2005)\citenamefont{{G{\'o}rski},
  {Hivon}, {Banday}, {Wandelt}, {Hansen}, {Reinecke}, and
  {Bartelmann}}}]{healpix}
\bibinfo{author}{\bibfnamefont{K.~M.} \bibnamefont{{G{\'o}rski}}},
  \bibinfo{author}{\bibfnamefont{E.}~\bibnamefont{{Hivon}}},
  \bibinfo{author}{\bibfnamefont{A.~J.} \bibnamefont{{Banday}}},
  \bibinfo{author}{\bibfnamefont{B.~D.} \bibnamefont{{Wandelt}}},
  \bibinfo{author}{\bibfnamefont{F.~K.} \bibnamefont{{Hansen}}},
  \bibinfo{author}{\bibfnamefont{M.}~\bibnamefont{{Reinecke}}},
  \bibnamefont{and}
  \bibinfo{author}{\bibfnamefont{M.}~\bibnamefont{{Bartelmann}}},
  \bibinfo{journal}{\apj} \textbf{\bibinfo{volume}{622}}, \bibinfo{pages}{759}
  (\bibinfo{year}{2005}), \eprint{astro-ph/0409513}.

\bibitem[{\citenamefont{{Stompor} et~al.}(2009)\citenamefont{{Stompor},
  {Leach}, {Stivoli}, and {Baccigalupi}}}]{Stompor2009}
\bibinfo{author}{\bibfnamefont{R.}~\bibnamefont{{Stompor}}},
  \bibinfo{author}{\bibfnamefont{S.}~\bibnamefont{{Leach}}},
  \bibinfo{author}{\bibfnamefont{F.}~\bibnamefont{{Stivoli}}},
  \bibnamefont{and}
  \bibinfo{author}{\bibfnamefont{C.}~\bibnamefont{{Baccigalupi}}},
  \bibinfo{journal}{\mnras} \textbf{\bibinfo{volume}{392}},
  \bibinfo{pages}{216} (\bibinfo{year}{2009}), \eprint{0804.2645}.

\bibitem[{\citenamefont{{Errard} et~al.}(2011)\citenamefont{{Errard},
  {Stivoli}, and {Stompor}}}]{Errard2011}
\bibinfo{author}{\bibfnamefont{J.}~\bibnamefont{{Errard}}},
  \bibinfo{author}{\bibfnamefont{F.}~\bibnamefont{{Stivoli}}},
  \bibnamefont{and}
  \bibinfo{author}{\bibfnamefont{R.}~\bibnamefont{{Stompor}}},
  \bibinfo{journal}{\prd} \textbf{\bibinfo{volume}{84}}, \bibinfo{eid}{063005}
  (\bibinfo{year}{2011}).

\bibitem[{\citenamefont{{Thorne} et~al.}(2017)\citenamefont{{Thorne},
  {Dunkley}, {Alonso}, and {N{\ae}ss}}}]{2017MNRAS.469.2821T}
\bibinfo{author}{\bibfnamefont{B.}~\bibnamefont{{Thorne}}},
  \bibinfo{author}{\bibfnamefont{J.}~\bibnamefont{{Dunkley}}},
  \bibinfo{author}{\bibfnamefont{D.}~\bibnamefont{{Alonso}}}, \bibnamefont{and}
  \bibinfo{author}{\bibfnamefont{S.}~\bibnamefont{{N{\ae}ss}}},
  \bibinfo{journal}{\mnras} \textbf{\bibinfo{volume}{469}},
  \bibinfo{pages}{2821} (\bibinfo{year}{2017}), \eprint{1608.02841}.

\bibitem[{\citenamefont{{Tegmark} et~al.}(1997)\citenamefont{{Tegmark},
  {Taylor}, and {Heavens}}}]{Tegmark1997}
\bibinfo{author}{\bibfnamefont{M.}~\bibnamefont{{Tegmark}}},
  \bibinfo{author}{\bibfnamefont{A.~N.} \bibnamefont{{Taylor}}},
  \bibnamefont{and} \bibinfo{author}{\bibfnamefont{A.~F.}
  \bibnamefont{{Heavens}}}, \bibinfo{journal}{\apj}
  \textbf{\bibinfo{volume}{480}}, \bibinfo{pages}{22} (\bibinfo{year}{1997}),
  \eprint{astro-ph/9603021}.

\bibitem[{\citenamefont{{Ward} et~al.}(2018)\citenamefont{{Ward}, {Alonso},
  {Errard}, {Devlin}, and {Hasselfield}}}]{2018arXiv180307630W}
\bibinfo{author}{\bibfnamefont{J.~T.} \bibnamefont{{Ward}}},
  \bibinfo{author}{\bibfnamefont{D.}~\bibnamefont{{Alonso}}},
  \bibinfo{author}{\bibfnamefont{J.}~\bibnamefont{{Errard}}},
  \bibinfo{author}{\bibfnamefont{M.~J.} \bibnamefont{{Devlin}}},
  \bibnamefont{and}
  \bibinfo{author}{\bibfnamefont{M.}~\bibnamefont{{Hasselfield}}},
  \bibinfo{journal}{ArXiv e-prints}  (\bibinfo{year}{2018}),
  \eprint{1803.07630}.

\bibitem[{\citenamefont{{Thuong Hoang} et~al.}(2017)\citenamefont{{Thuong
  Hoang}, {Patanchon}, {Bucher}, {Matsumura}, {Banerji}, {Ishino}, {Hazumi},
  and {Delabrouille}}}]{2017JCAP...12..015T}
\bibinfo{author}{\bibfnamefont{D.}~\bibnamefont{{Thuong Hoang}}},
  \bibinfo{author}{\bibfnamefont{G.}~\bibnamefont{{Patanchon}}},
  \bibinfo{author}{\bibfnamefont{M.}~\bibnamefont{{Bucher}}},
  \bibinfo{author}{\bibfnamefont{T.}~\bibnamefont{{Matsumura}}},
  \bibinfo{author}{\bibfnamefont{R.}~\bibnamefont{{Banerji}}},
  \bibinfo{author}{\bibfnamefont{H.}~\bibnamefont{{Ishino}}},
  \bibinfo{author}{\bibfnamefont{M.}~\bibnamefont{{Hazumi}}}, \bibnamefont{and}
  \bibinfo{author}{\bibfnamefont{J.}~\bibnamefont{{Delabrouille}}},
  \bibinfo{journal}{\jcap} \textbf{\bibinfo{volume}{12}}, \bibinfo{eid}{015}
  (\bibinfo{year}{2017}), \eprint{1706.09486}.

\bibitem[{\citenamefont{{Errard} and {Stompor}}(2012)}]{Errard2012}
\bibinfo{author}{\bibfnamefont{J.}~\bibnamefont{{Errard}}} \bibnamefont{and}
  \bibinfo{author}{\bibfnamefont{R.}~\bibnamefont{{Stompor}}},
  \bibinfo{journal}{\prd} \textbf{\bibinfo{volume}{85}}, \bibinfo{eid}{083006}
  (\bibinfo{year}{2012}), \eprint{1203.5285}.

\bibitem[{\citenamefont{{Planck Collaboration}
  et~al.}(2016{\natexlab{b}})\citenamefont{{Planck Collaboration}, {Ade},
  {Aghanim}, {Arnaud}, {Ashdown}, {Aumont}, {Baccigalupi}, {Banday},
  {Barreiro}, {Bartlett} et~al.}}]{2016A&A...594A..13P}
\bibinfo{author}{\bibnamefont{{Planck Collaboration}}},
  \bibinfo{author}{\bibfnamefont{P.~A.~R.} \bibnamefont{{Ade}}},
  \bibinfo{author}{\bibfnamefont{N.}~\bibnamefont{{Aghanim}}},
  \bibinfo{author}{\bibfnamefont{M.}~\bibnamefont{{Arnaud}}},
  \bibinfo{author}{\bibfnamefont{M.}~\bibnamefont{{Ashdown}}},
  \bibinfo{author}{\bibfnamefont{J.}~\bibnamefont{{Aumont}}},
  \bibinfo{author}{\bibfnamefont{C.}~\bibnamefont{{Baccigalupi}}},
  \bibinfo{author}{\bibfnamefont{A.~J.} \bibnamefont{{Banday}}},
  \bibinfo{author}{\bibfnamefont{R.~B.} \bibnamefont{{Barreiro}}},
  \bibinfo{author}{\bibfnamefont{J.~G.} \bibnamefont{{Bartlett}}},
  \bibnamefont{et~al.}, \bibinfo{journal}{\aap} \textbf{\bibinfo{volume}{594}},
  \bibinfo{eid}{A13} (\bibinfo{year}{2016}{\natexlab{b}}), \eprint{1502.01589}.

\bibitem[{\citenamefont{{Planck Collaboration}
  et~al.}(2016{\natexlab{c}})\citenamefont{{Planck Collaboration}, {Adam},
  {Ade}, {Aghanim}, {Arnaud}, {Aumont}, {Baccigalupi}, {Banday}, {Barreiro},
  {Bartlett} et~al.}}]{Planck_XXX_2016}
\bibinfo{author}{\bibnamefont{{Planck Collaboration}}},
  \bibinfo{author}{\bibfnamefont{R.}~\bibnamefont{{Adam}}},
  \bibinfo{author}{\bibfnamefont{P.~A.~R.} \bibnamefont{{Ade}}},
  \bibinfo{author}{\bibfnamefont{N.}~\bibnamefont{{Aghanim}}},
  \bibinfo{author}{\bibfnamefont{M.}~\bibnamefont{{Arnaud}}},
  \bibinfo{author}{\bibfnamefont{J.}~\bibnamefont{{Aumont}}},
  \bibinfo{author}{\bibfnamefont{C.}~\bibnamefont{{Baccigalupi}}},
  \bibinfo{author}{\bibfnamefont{A.~J.} \bibnamefont{{Banday}}},
  \bibinfo{author}{\bibfnamefont{R.~B.} \bibnamefont{{Barreiro}}},
  \bibinfo{author}{\bibfnamefont{J.~G.} \bibnamefont{{Bartlett}}},
  \bibnamefont{et~al.}, \bibinfo{journal}{\aap} \textbf{\bibinfo{volume}{586}},
  \bibinfo{eid}{A133} (\bibinfo{year}{2016}{\natexlab{c}}), \eprint{1409.5738}.

\bibitem[{\citenamefont{{Herv{\'{\i}}as-Caimapo}
  et~al.}(2017)\citenamefont{{Herv{\'{\i}}as-Caimapo}, {Bonaldi}, and
  {Brown}}}]{2017MNRAS.468.4408H}
\bibinfo{author}{\bibfnamefont{C.}~\bibnamefont{{Herv{\'{\i}}as-Caimapo}}},
  \bibinfo{author}{\bibfnamefont{A.}~\bibnamefont{{Bonaldi}}},
  \bibnamefont{and} \bibinfo{author}{\bibfnamefont{M.~L.}
  \bibnamefont{{Brown}}}, \bibinfo{journal}{\mnras}
  \textbf{\bibinfo{volume}{468}}, \bibinfo{pages}{4408} (\bibinfo{year}{2017}),
  \eprint{1701.02277}.

\bibitem[{\citenamefont{{Kogut}}(2012)}]{2012ApJ...753..110K}
\bibinfo{author}{\bibfnamefont{A.}~\bibnamefont{{Kogut}}},
  \bibinfo{journal}{\apj} \textbf{\bibinfo{volume}{753}}, \bibinfo{eid}{110}
  (\bibinfo{year}{2012}), \eprint{1205.4041}.

\bibitem[{\citenamefont{{Rotti} and
  {Huffenberger}}(2016)}]{2016JCAP...09..034R}
\bibinfo{author}{\bibfnamefont{A.}~\bibnamefont{{Rotti}}} \bibnamefont{and}
  \bibinfo{author}{\bibfnamefont{K.}~\bibnamefont{{Huffenberger}}},
  \bibinfo{journal}{\jcap} \textbf{\bibinfo{volume}{9}}, \bibinfo{eid}{034}
  (\bibinfo{year}{2016}), \eprint{1604.08946}.

\bibitem[{\citenamefont{Lewis et~al.}(2000)\citenamefont{Lewis, Challinor, and
  Lasenby}}]{camb-1}
\bibinfo{author}{\bibfnamefont{A.}~\bibnamefont{Lewis}},
  \bibinfo{author}{\bibfnamefont{A.}~\bibnamefont{Challinor}},
  \bibnamefont{and} \bibinfo{author}{\bibfnamefont{A.}~\bibnamefont{Lasenby}},
  \bibinfo{journal}{Astrophys. J.} \textbf{\bibinfo{volume}{538}},
  \bibinfo{pages}{473} (\bibinfo{year}{2000}), \eprint{astro-ph/9911177}.

\bibitem[{\citenamefont{Howlett et~al.}(2012)\citenamefont{Howlett, Lewis,
  Hall, and Challinor}}]{camb-2}
\bibinfo{author}{\bibfnamefont{C.}~\bibnamefont{Howlett}},
  \bibinfo{author}{\bibfnamefont{A.}~\bibnamefont{Lewis}},
  \bibinfo{author}{\bibfnamefont{A.}~\bibnamefont{Hall}}, \bibnamefont{and}
  \bibinfo{author}{\bibfnamefont{A.}~\bibnamefont{Challinor}},
  \bibinfo{journal}{JCAP} \textbf{\bibinfo{volume}{1204}}, \bibinfo{pages}{027}
  (\bibinfo{year}{2012}), \eprint{1201.3654}.

\end{thebibliography}

\end{document}